\documentclass{aa}  
\usepackage{lscape}
\usepackage{listings}
\usepackage{natbib}
\bibpunct{(}{)}{;}{a}{}{,} 
\usepackage{booktabs,caption}
\usepackage[flushleft]{threeparttable}
\usepackage{adjustbox}

\newcommand\gaia{\textit{Gaia}}
\newcommand{\gdr}[1]{\gaia~DR#1}
\newcommand\muas{\ensuremath{\mu\text{as}}}
\newcommand\masyr{\ensuremath{\text{mas~yr}^{-1}}}

\newcommand\kms{\ensuremath{\text{km~s}^{-1}}}

\newcommand\gbp{\ensuremath{G_\mathrm{BP}}}
\newcommand\grp{\ensuremath{G_\mathrm{RP}}}
\newcommand\bpminrp{\ensuremath{(\gbp-\grp)}}
\newcommand\vlos{\ensuremath{v_\mathrm{los}}}
\newcommand\pmra{\ensuremath{\mu_{\alpha*}}}
\newcommand\pmdec{\ensuremath{\mu_{\delta}}}
\newcommand\feh{\ensuremath{[\mathrm{Fe}/\mathrm{H}]}}

\begin{document}

\title{A 3D view of dwarf galaxies with \gaia\ and VLT/FLAMES\\I. The Sculptor dwarf spheroidal\thanks{Tables E.1 and E.2 are only available online via the CDS.}$^,$\thanks{Based on VLT/FLAMES observations collected at the European Organisation for Astronomical Research (ESO) in the Southern Hemisphere under programmes 171.B-0588, 072.D-0245, 076.B-0391, 079.B-0435, 593.D-0309, 098.D-0160, 0100.B-0337, 0101.D-0210, 0101.B-0189, and 0102.B-0786.} }

\titlerunning{A 3D view of Sculptor}
\authorrunning{ Tolstoy, Sk\'ulad\'ottir, Battaglia et al.}
\author{Eline Tolstoy, \inst{1} \'Asa Sk\'ulad\'ottir, \inst{2,3} Giuseppina Battaglia, \inst{4, 5} Anthony G.A. Brown, \inst{6} Davide Massari, \inst{7, 1}  \\
Michael J. Irwin, \inst{8} Else Starkenburg, \inst{1} Stefania Salvadori, \inst{2,3} Vanessa Hill, \inst{9} Pascale Jablonka, \inst{10,11}\\ Maurizio Salaris, \inst{12} Thom van Essen, \inst{1} Carla Olsthoorn, \inst{1} Amina Helmi \inst{1} \&
John Pritchard \inst{13}
}
\institute{Kapteyn Astronomical Institute, University of Groningen, PO Box 800,
             9700AV Groningen, the Netherlands\\ \email{etolstoy@astro.rug.nl}
             \and
             Dipartimento di Fisica e Astronomia, Universit\'a degli Studi di Firenze, Via G. Sansone 1, I-50019 Sesto Fiorentino, Italy 
             \and
             INAF/Osservatorio Astrofisico di Arcetri, Largo E. Fermi 5, I-50125 Firenze, Italy
             \and
             Instituto de Astrof\'isica de Canarias, Calle V\'ia L\'actea s/n, E-38206 La Laguna, Tenerife, Spain
             \and 
             Universidad de La Laguna, Avda. Astrof\'isico Fco. S\'anchez, E-38205 La Laguna, Tenerife, Spain
             \and
             Leiden Observatory, Leiden University, Niels Bohrweg 2, NL-2333 CA Leiden, the Netherlands
             \and
             INAF - Osservatorio di Astrofisica e Scienza dello Spazio di Bologna, Via Gobetti 93/3, I-40129 Bologna, Italy
             \and
             Institute of Astronomy, Madingley Road, Cambridge CB3 0HA, UK
            \and
             Laboratoire Lagrange, Universit\'e de Nice Sophia Antipolis, CNRS, Observatoire de la C\^ote d'Azur, CS34229, F-06304 Nice Cedex 4, France
             \and 
             Institute of Physics, Laboratory of Astrophysics, Ecole Polytechnique F\'ed\'erale de Lausanne (EPFL), Observatoire de Sauverny, 1290, Versoix, Switzerland; 
             \and
             GEPI, Observatoire de Paris, CNRS, Universit\'e Paris Diderot, 92125, Meudon Cedex, France
             \and
             Astrophysics Research Institute, Liverpool John Moores University, 146 Brownlow Hill, Liverpool L3 5RF, UK;
             \and
             European Southern Observatory, Karl-Schwarzschild-Str. 2, D-85748 Garching bei M\"unchen, Germany
             }

\abstract{
We present a new homogeneous survey of VLT/FLAMES LR8 line-of-sight radial velocities (\vlos) for 1604 resolved red giant branch stars in the Sculptor dwarf spheroidal galaxy. In addition, we provide reliable Ca~II triplet metallicities, \feh, for 1339 of these stars. From this combination of new observations (2257 individual spectra) with ESO archival data (2389 spectra), we obtain the largest and most complete sample of \vlos\ and \feh\ measurements for individual stars in any dwarf galaxy. Our sample includes VLT/FLAMES LR8 spectra for $\sim$55\% of the red giant branch stars at G $< 20$ from \gaia\ DR3, and $>70$\% of the brightest stars, G $<18.75$. Our spectroscopic velocities are combined with \gaia\ DR3 proper motions and parallax measurements for a new and more precise membership analysis. We look again at the global characteristics of Sculptor, deriving a mean metallicity of $\langle$\feh$\rangle = -1.82\pm0.45$ and a mean line-of-sight velocity of $\langle$\vlos$\rangle = +111.2\pm0.25$km/s. There is a clear metallicity gradient in Sculptor, $-0.7$deg/dex, with the most metal-rich population being the most centrally concentrated. Furthermore, the most metal-poor population in Sculptor, \feh~$< -2.5$, appears to show kinematic properties distinct from the rest of the stellar population. Finally, we combine our results with the exquisite \textit{Gaia} DR3 multi-colour photometry to further investigate the colour-magnitude diagram of the resolved stellar population in Sculptor. Our detailed analysis shows a  similar global picture as previous studies, but with much more precise detail, revealing that Sculptor has more complex properties than previously thought. This survey emphasises the role of the stellar spectroscopy technique and this galaxy as a benchmark system for modelling galaxy formation and evolution on small scales.
}


 \maketitle

\section{Introduction}

Classical dwarf spheroidals (dSphs) such as Sculptor occupy the faint end of the galaxy luminosity function. In the immediate vicinity of the Milky Way, they are fully resolved star by star down to luminosities well below the oldest main sequence turnoff. As apparently strongly dark-matter-dominated systems, they are powerful test beds for models of dark matter, as well as the processes that drive star formation and chemical enrichment at early times. Compared to larger and more massive galaxies, dwarf galaxies are much more sensitive to the energetic process of star formation and stellar feedback. Thus, they allow us to distinguish what is universal in galaxy evolution from what are more stochastic processes at work in small systems. We are able to study dSphs in detail because the total number of stars is large enough that their dynamic and chemical properties can be traced using many hundreds of individual red giant branch (RGB) stars, which can be accurately compared to detailed predictions from models and simulations. 

The \gaia\ mission \citep{Prusti16, Vallenari22} has provided a breakthrough in our understanding of resolved stellar populations in and around the Milky Way, and even in nearby dwarf galaxies. \gaia\ has no equal when it comes to astrometry, which, combined with its exquisite photometry, is a uniquely powerful tool for accurately selecting and analysing member stars throughout nearby resolved dwarf galaxies, especially in the sparse outskirts. However, \gaia\ will not provide complete line-of-sight (los) velocities, \vlos, or metallicities for individual stars in most of the  satellite dwarf galaxies. This provides a clear motivation to create the most complete and homogeneous survey of 3D velocities and metallicities of RGB stars in dSphs down to the magnitude limit that matches \gaia\ astrometry. This allows accurate comparisons with 3D dynamical models and in turn allows us to look for rotation and tidal effects and determine the relation with the orbit around the Milky Way, while also providing a more complete understanding of the chemical evolution over time within these systems.

The Sculptor dSph is a satellite of the Milky Way at a distance of 83.9\,kpc \citep{MV15}.\ It is a key test case for models of galaxy formation and evolution as it is a metal-poor fossil from the early Universe. It stopped forming stars $\sim$8-10\,Gyr ago \citep{deBoer12, Bettinelli19, dlR22}. Thus it provides a clear vision of ancient times, unobscured by recent star formation episodes. Sculptor has been studied extensively in relation to: its star formation history (SFH; e.g.~\citealt{Monk99,Dolphin02,deBoer12,Weisz14,Savino18,Bettinelli19}); the chemical abundances of its stellar population \citep[e.g.][]{NorrisBessell78, Shetrone98, Shetrone03, Tolstoy03, Geisler05, Clementini05, Kirby09,Kirby19,Starkenburg13, Skuladottir15b,Skuladottir15a,Skuladottir17,Skuladottir18,Skuladottir19,Skuladottir20,Skuladottir21,Jablonka15,Simon15,Lardo16,Chiti18,  Hill19, Salgado19,Reichert20,Tang22}; and its kinematics \citep[e.g.][]{Queloz95, Tolstoy01, Tolstoy04,Battaglia08b,Walker09,Zhu16, Massari18, Iorio19}. In addition, the RR~Lyrae variable stars in the Sculptor dSph have long been studied \citep[e.g.][]{BaadeHubble39, Kaluzny95, MV15}.

Despite the simplicity of Sculptor's SFH, and its spheroidal structure, its kinematics are surprisingly complex. Previous studies have revealed at least two distinct components, distinguished by different metallicity distributions, spatial distributions, and kinematics \citep[e.g.][]{Tolstoy04,Battaglia08b, WalkPenn11, Breddels13, BreHel13}. By determining the SFH of Sculptor at different radii, \cite{deBoer12} showed that older, more metal-poor populations are present throughout the galaxy, while younger, more metal-rich stars tend to be more centrally concentrated, confirming what was also shown by other{ tracers of specific stellar populations} \citep[e.g.][]{HK99, Harbeck01, Tolstoy04}. These observational results mean that it is not straightforward to model the properties of the Sculptor dSph and uniquely determine the evolutionary path that led to the galaxy we see today \citep[e.g.][]{Battaglia08a,Salvadori08, Salvadori15, Lokas09, Revaz09, AgnelloEvans12, RomanoStark13,Zhu16,Strigari17,dlR22}.

The existing spectroscopic datasets of Sculptor provide metallicities (\feh) and los velocities (v$_{los}$) that are neither complete nor homogeneous.  The  incompleteness is typically higher at faint magnitudes as well as in the outer regions that are dominated by the most ancient and metal-poor stars. Due to the metallicity gradient in Sculptor, it is not possible to simply to correct for the incompleteness in the outer regions using the results from the inner \citep[e.g.][]{RomanoStark13}. A complete and unbiased set of measurements of the \feh\ and v$_{los}$ distributions over the entire RGB age range in Sculptor is needed to accurately constrain both the buildup of metals as a result of extended star formation and the loss of metals due to stellar feedback \citep[e.g.][]{Dekel86, GilWyse91,FerraraTolstoy00,LanfranchiMatteucci04,Fenner06, Marcolini08, SalvadoriFerrara09, Revaz09, Revaz18}. A large and complete sample of RGB stars is also important because the quantification of scatter in the properties of individual stars is critical for a correct understanding of galactic-scale dynamics \citep[e.g.][]{Battaglia13} and chemical evolution \citep[e.g.][]{Cayrel04}.

\begin{figure}[t]
\centering
\includegraphics[width=1\linewidth]{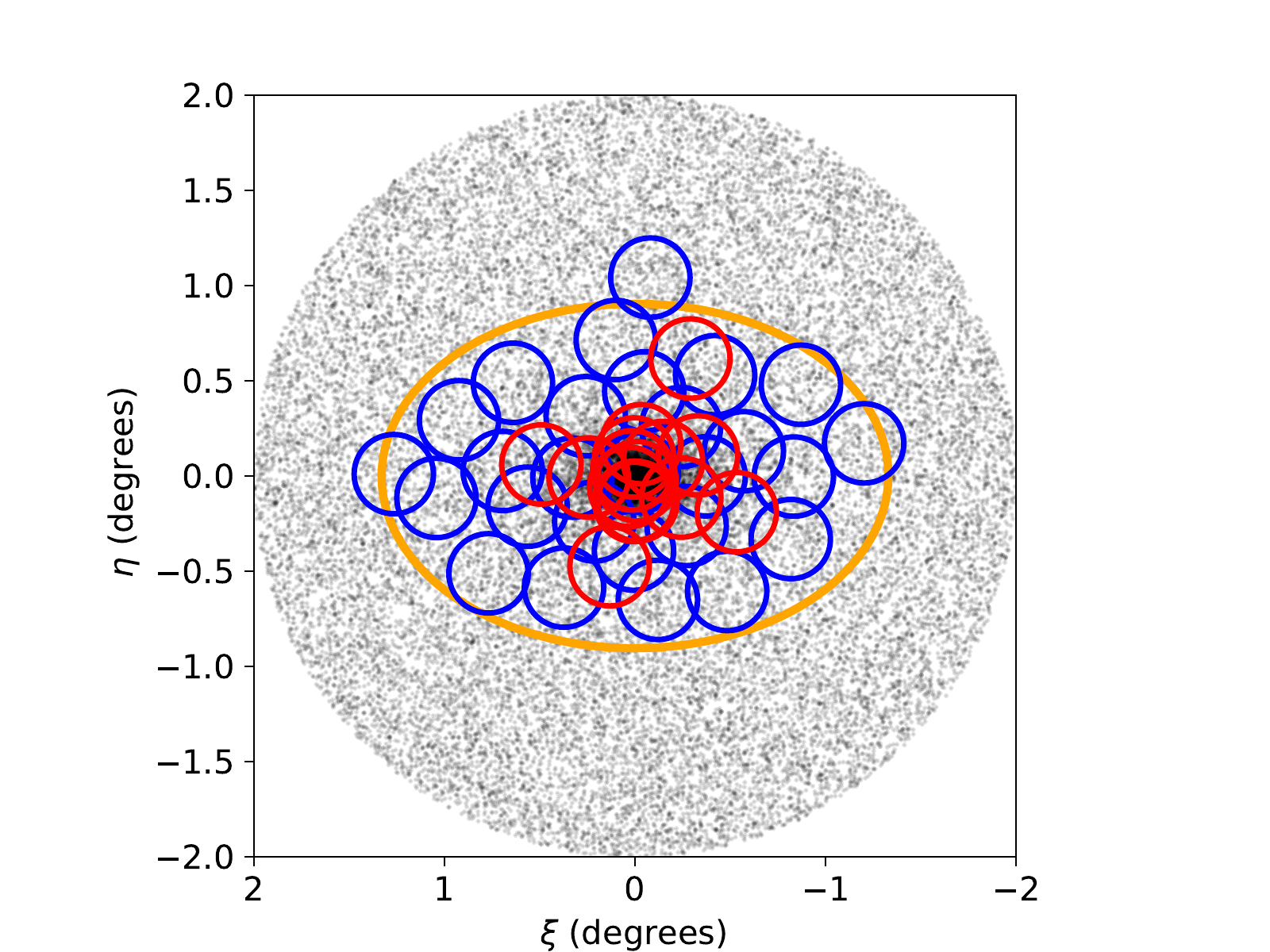}
\caption{ \gdr{3} catalogue (black dots) within 2$^\circ$ of the centre of the Sculptor dSph, consisting of 36\,185 objects that possess photometry, proper motion, and parallax measurements. Positions of the 44 available VLT/FLAMES LR8 fields ($\sim25^\prime$ diameter), many of which overlap, are shown with circles. In blue are the 27 previously analysed fields, and in red the 17 new fields. The nominal tidal radius of Sculptor { \citep[from][]{Irwin95}} is shown as an orange ellipse.}
\label{fig1a}
\end{figure}

A long-standing challenge for this kind of study has been the accurate determination of the membership of individual stars (e.g. \citealt{BattStar12}). The \gaia\ proper motions and parallaxes are relatively new, first provided in \textit{Gaia} Data Release (DR) 2 \citep{Brown18}, and are regularly being improved, most recently in \gdr{3}  \citep{Brown21}. This highly accurate astrometry, which uses detailed, high-precision measurements of  individual stars, provides unique insights into the global properties of nearby resolved galaxies  \citep[e.g.][]{Helmi18, McV20, Battaglia22}. Thus we take the opportunity to re-examine the Very Large Telescope (VLT) Fibre Large Array Multi Element Spectrograph (FLAMES) Ca~II triplet (CaT) surveys that cover the whole area of the Sculptor dSph on the sky (see Fig.~\ref{fig1a}) and combine them with the \gdr{3} catalogue to create a uniquely detailed and complete catalogue.
Including both new and reprocessed archival VLT/FLAMES data, we present the most complete and homogeneous survey of the 3D velocities and metallicities of the resolved stellar population in any galaxy to date, enhancing Sculptor as a benchmark system for understanding galaxy formation and evolution on small scales.

\section{VLT/FLAMES spectroscopy}

The European Southern Observatory (ESO) VLT FLAMES spectrograph in GIRAFFE mode using the LR8 grating has provided spectra of RGB stars in the Sculptor dSph galaxy over nearly 20 years. Here we present extensive recent observations together with consistently analysed older programmes. All observations, recent and from the distant past, have been processed uniformly with the most up to date pipeline and additional software to give self-consistent velocities and metallicities, with accompanying errors, across all VLT/FLAMES LR8 datasets. 

\begin{figure*}[ht]
\includegraphics[width=1\linewidth]{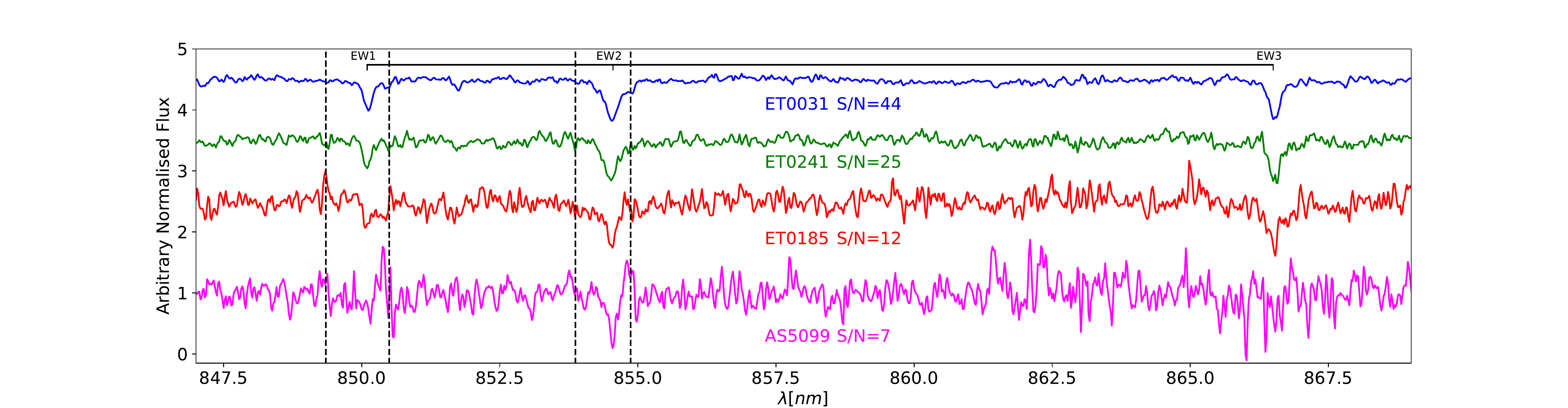}
\captionof{figure}{{ Normalised spectra of four RGB stars in the Sculptor dSph galaxy, all with \feh$\sim -1.7$, for a range of S/N. The three CaT absorption lines for stars in the Sculptor dSph are marked at $\lambda\lambda$8500\,\AA (EW1), 8545\,\AA (EW2), and 8665\,\AA (EW3). Vertical dashed black lines show the central positions of strong sky lines that can cause problematic residuals in low S/N spectra.}}
\label{specsn}
\end{figure*}

\begin{figure*}
\includegraphics[width=1\linewidth]{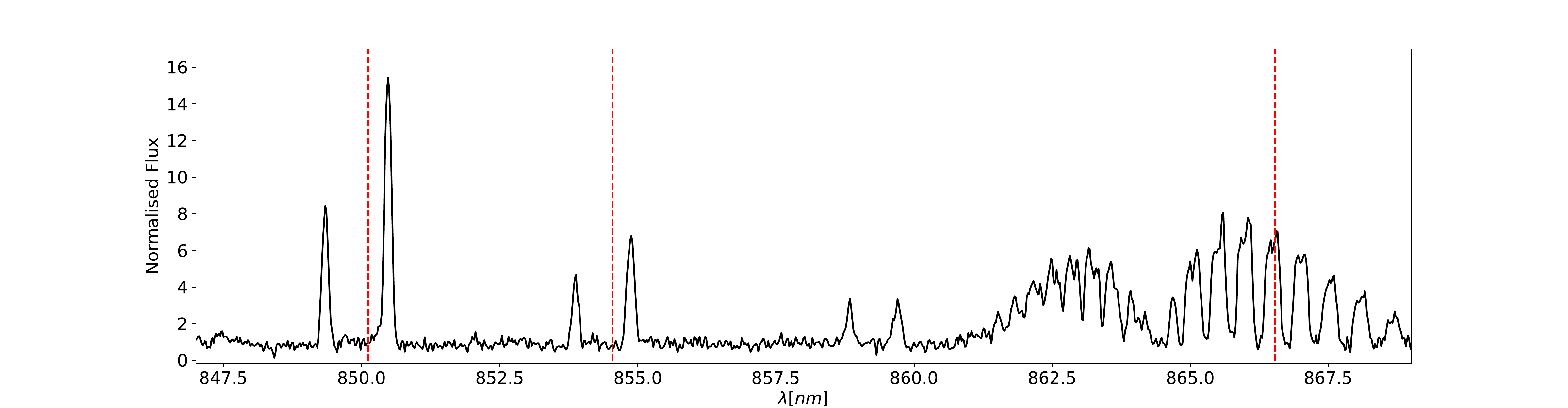}
\captionof{figure}{{ Normalised sky spectrum from a single sky fibre, from the same observation and with the same normalisation 
as the top three spectra in Fig.~\ref{specsn}. The central positions of the three CaT absorption lines at the mean velocity of the Sculptor dSph are marked as dashed red lines.}}
\label{specsky}
\end{figure*}

\begin{figure*}
\includegraphics[width=1\linewidth]{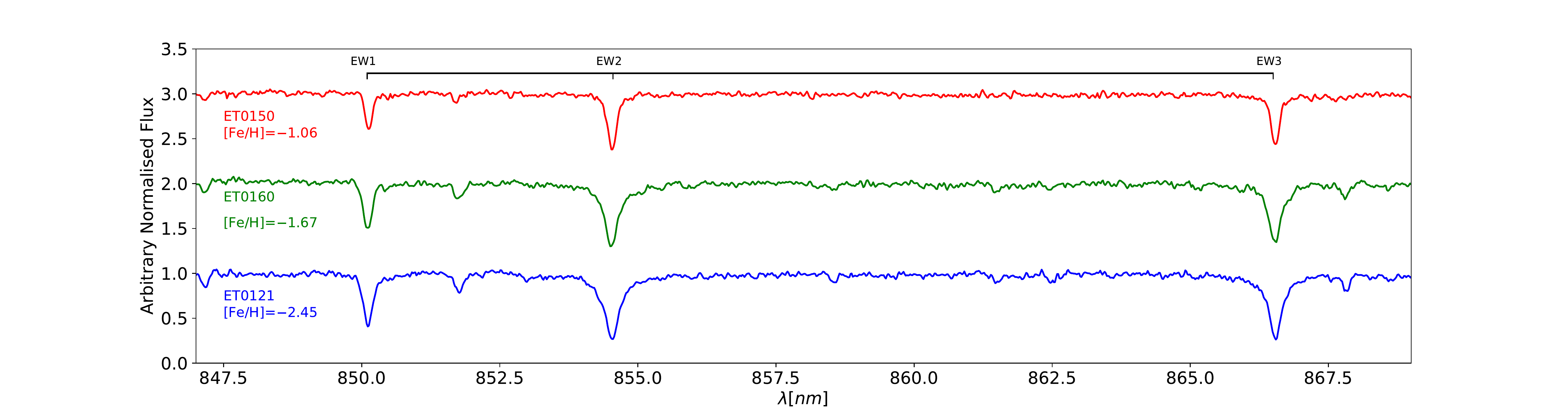}
\captionof{figure}{{ Normalised high S/N ($\sim65-75$) spectra of three RGB stars in the Sculptor dSph galaxy, with a range of metallicities, \feh. The three CaT absorption lines are marked.} }
\label{specfeh}
\end{figure*}

\subsection{VLT/FLAMES LR8 observations}

We  extracted all spectra of individual stars in the direction of the Sculptor dSph galaxy from the ESO VLT archive observed with the FLAMES/GIRAFFE LR8 grating, covering the CaT spectral region ($\lambda\lambda$8498, 8542, 8662\,\AA). This includes 67 independent observations from 10 different observation programmes\footnote{171.B-0588 PI: Tolstoy (DART LP); 072.D-0245 PI: Hill (FLAMES GTO); 076.B-0391 PI: Battaglia; 079.B-0435 PI: Battaglia; 593.D-0309 PI: Battaglia; 098.D-0160 PI: de Boer; 0100.B-0337 PI: Tolstoy; 0101.D-0210  PI: Sk\'{u}lad\'{o}ttir;  0101.B-0189  PI: Tolstoy; 0102.B-0786  PI: Tolstoy}, taken between 2003 and 2018 (see Appendix~\ref{app:results}, Table~\ref{table1}), of 44 different pointings, 17 of which are new (see Fig.~\ref{fig1a}). All the observations from 2007 and earlier have been previously presented (e.g. ~\citealt{Tolstoy04, Battaglia08b, Battaglia08a, Starkenburg10}) and have also provided the basis of searches for extremely metal-poor stars in Sculptor (e.g. ~\citealt{Tafelmeyer10, Starkenburg13, Jablonka15}). The VLT/FLAMES LR8 observations are predominantly concentrated in the central region of the Sculptor dSph, where most of the stars are to be found. It also extends out to the nominal tidal radius (see Fig.~\ref{fig1a}), as defined by \citet{Irwin95}. 
Irwin \& Hatzidimitriou determined the tidal radius by fitting a King profile to the number density surface  profile of stars along the los to Sculptor, taking into account the presence of Galactic foreground/background. The ellipticity was found to be e=0.32 $\pm$ 0.03 and the position angle pa=99 $\pm$ 1~degrees.

The large collection of new VLT/FLAMES LR8 observations also includes repeat observations monitoring the same targets in the central field. Upcoming work  (Arroyo Polonio et al., in prep.) will use these data for a full analysis of the properties of binary stars in the Sculptor dSph.

\subsection{Processing the VLT/FLAMES LR8 spectra}

The first step in producing a uniform catalogue of \vlos\ and \feh\ measurements for RGB stars is to (re-)process all the available VLT/FLAMES LR8 observations with the most recent ESO pipeline (via the esoreflex tool; ~\citealt{Freudling13}), including cosmic ray removal, heliocentric correction, { and error arrays for each extracted spectrum}. This results in { 4646} individual spectra, of which{ 2257} are new (numbers per pointing are given in Table~\ref{table1}). The total number includes multiple measurements of the same stars, and also Galactic foreground stars and spectra with too low signal-to-noise ratio (S/N$<$7) to be consistently reliable. These are all removed from further analysis, as described in Sect.~\ref{sec:mem}

The ESO VLT/FLAMES data reduction pipeline provides calibrated spectra with the instrumental effects removed, as well as new statistical error arrays for each extracted spectrum, which were not available when the older data were analysed in 2008. These are the standard deviation of the re-sampled fluxes for each wavelength bin. The additional processing steps, still required after the ESO pipeline, are carried out using software developed over many years by Mike Irwin. An earlier version of this software was described in \citet{Battaglia08a}. The software used here has  been adapted to make use of the error arrays provided by the ESO pipeline. This improves the reliability of the error determinations on the \vlos\ and \feh\  measurements. Individual Fortran programmes are used, starting with adding accurate, well calibrated \gdr{3} photometry converted to V magnitude \citep{Riello21}, for each observed star into the fibre fits tables of the pipeline output. This photometry is later used to ensure uniform results in determining the CaT metallicities. All stars observed with VLT/FLAMES are found in the \gaia\ catalogue except one (AS5360\footnote{ \gdr{3} ID: 5003198370796137728; RA: 00 59 48.55; Dec.: $-$33 48 14; G$\sim 19.77$. From VLT/FLAMES: $\vlos = 123$ km/s; \feh$ \sim -1.4$}), which is in the \gdr{3} photometric catalogue, but it lacks astrometric measurements. This single star, although it is a likely member based on the VLT/FLAMES spectroscopy, was removed from further consideration.

Before sky subtraction a simple check was made for each VLT/FLAMES field that the ESO pipeline wavelength calibration, which uses the internal arcs taken in daytime, matches the known positions of the numerous sky lines in the LR8 wavelength range. A global shift was then applied to all the spectra for each individual field to put all the fields on the same zero point. These shifts typically varied between 1 and 2.5 km/s, but there were also a few  observations with larger shifts. These shifts are listed in Table~\ref{table1}. When the velocity shifts become large this is most likely due to a problem with the guide star positioning in setting up the observation or the available day-time calibration frames. This global check is carried out to ensure that there are no systematic offsets in velocity between different VLT/FLAMES pointings. The individual spectra once corrected for these small velocity shifts were then sky subtracted, again using the dedicated software from Mike Irwin. The sample of 4646 spectra was then ready to be analysed to determine membership and spectroscopic \vlos\ and \feh. 

In Fig.~\ref{specsn} we show the effects of S/N on VLT/FLAMES LR8 spectra for stars with the same \feh, demonstrating how sky subtraction becomes more challenging at lower S/N. Figure~\ref{specsky} shows on the same scale as  Fig.~\ref{specsn} the strength of the sky lines compared to the CaT lines and the proximity of the strong sky lines. The dependence of the CaT line strength on metallicity is complex due to the variety of stellar atmospheric parameters ($T_\textsl{eff}$, $\log g$), as is most clearly illustrated in Fig.~1 of \citet{Starkenburg17}. We show three spectra from our VLT/FLAMES LR8 sample with varying metallicity in Fig.~\ref{specfeh}. It is clear that the wings of the lines change most going from high to low \feh. This also means that, especially at low S/N the higher metallicity stars are more affected by nearby sky lines as the CaT wings will tend to overlap with the sky line residuals. We try to remove as many spurious measurements coming from sky subtraction problems as we can unambiguously identify, even though they are few and have a very limited effect on the final results.

As the majority of the spectra available were taken before \gaia\ was launched, our early observations contain a significant number of non-members. This results in independent samples of velocities and proper motion selections, which allows more wide-ranging checks on the membership boundaries than might otherwise have been possible. From 2018 onwards we were able to make use of a priori information from the \gdr{2} proper motion catalogue, and as can be seen in Table~\ref{table1}, this dramatically increased the efficiency at picking out likely members of the Sculptor dSph for spectroscopic follow-up.  The multiple observations were used to test our error analysis more completely than has been possible in the past \citep[e.g.][and see our Appendix~\ref{app:lr8}]{Battaglia08a}. 

\subsection{Measuring velocities (\vlos ) and metallicities (\feh) from VLT/FLAMES LR8 spectra}

There has been extensive use of a variety of methods to determine the likely membership of stars in the Sculptor dSph: according to the spatial distribution \citep[e.g.][]{Demers80}, the colour-magnitude diagram (CMD) distribution \citep[e.g.][]{DaCosta84} and los velocities \citep[e.g.][]{ArmDaC86, Aar87, Queloz95} and more recently a combination of all of these, building up to the present day inclusion of parallax and proper motion from \gaia\ \citep[e.g.][]{Battaglia22}. There has also been extensive use of  metallicity indicators in low resolution spectra (R$<$8000), such as the CaT \citep[e.g.][]{Tolstoy01, Tolstoy04, Battaglia08a, Starkenburg10, Carrera13}, and the Mg~triplet \citep[e.g.][]{Walker09} to determine \feh and using  Ca H\&K lines as well as photometry \citep{Chiti18} and a broad spectral range \citep{Kirby10}.  Metallicities (and additional lines in the low resolution spectra) have also be used to refine membership determination of the stars in the Sculptor dSph \citep[e.g.][]{BattStar12}.

\subsubsection{The line-of-sight velocities (\vlos )}\label{sec:specvel}

The los velocities\footnote{The los velocities are always in the heliocentric system.}, \vlos, were determined from the VLT/FLAMES spectra using the three  CaT lines (Figs.~\ref{specsn} and \ref{specfeh}). In our processing we explored three different methods: (1)~a fit of the individual lines; (2)~cross correlation; and (3)~the maximum likelihood method (MLM). All three methods used a template RGB spectrum for comparison. All methods should (and did) produce very similar results; however, MLM was the most statistically robust, once the ESO pipeline error arrays were utilised, and it could  provide reliable error estimates. 
Significant differences in velocities determined for the same spectra with different methods ($\geq 5$km/s) were used to flag and remove (the few) clearly unreliable measurements at low S/N. This mismatch was typically created by poor sky subtraction at low S/N ($\lesssim 20$) where the presence of sky subtraction residuals created significant variations between different methods due to the proximity of strong sky lines to the CaT lines (Figs.~\ref{specsn} and \ref{specsky}). This effect can clearly be seen in the two lowest S/N spectra in Fig.~\ref{specsn}.
We tested the error determination using the sequence of repeated observations of the same stars (see Appendix~\ref{app:lr8}) for a subset of 96 stars from a monitoring programme looking for binary stars in Sculptor (Arroyo Polonio et al. in prep). It was immediately apparent that the S/N varies for the same stars over the programme and between observations (see Fig.~\ref{fig-sn}). This was most likely due to differences in fibre positioning between different acquisitions of the same field. There were also a few obviously spurious measurements that are likely due to imperfections in the spectra (e.g. problems with sky subtraction or continuum definition) that can lead to errors in the results.

\begin{figure}
\centering
\includegraphics[width=1\linewidth]{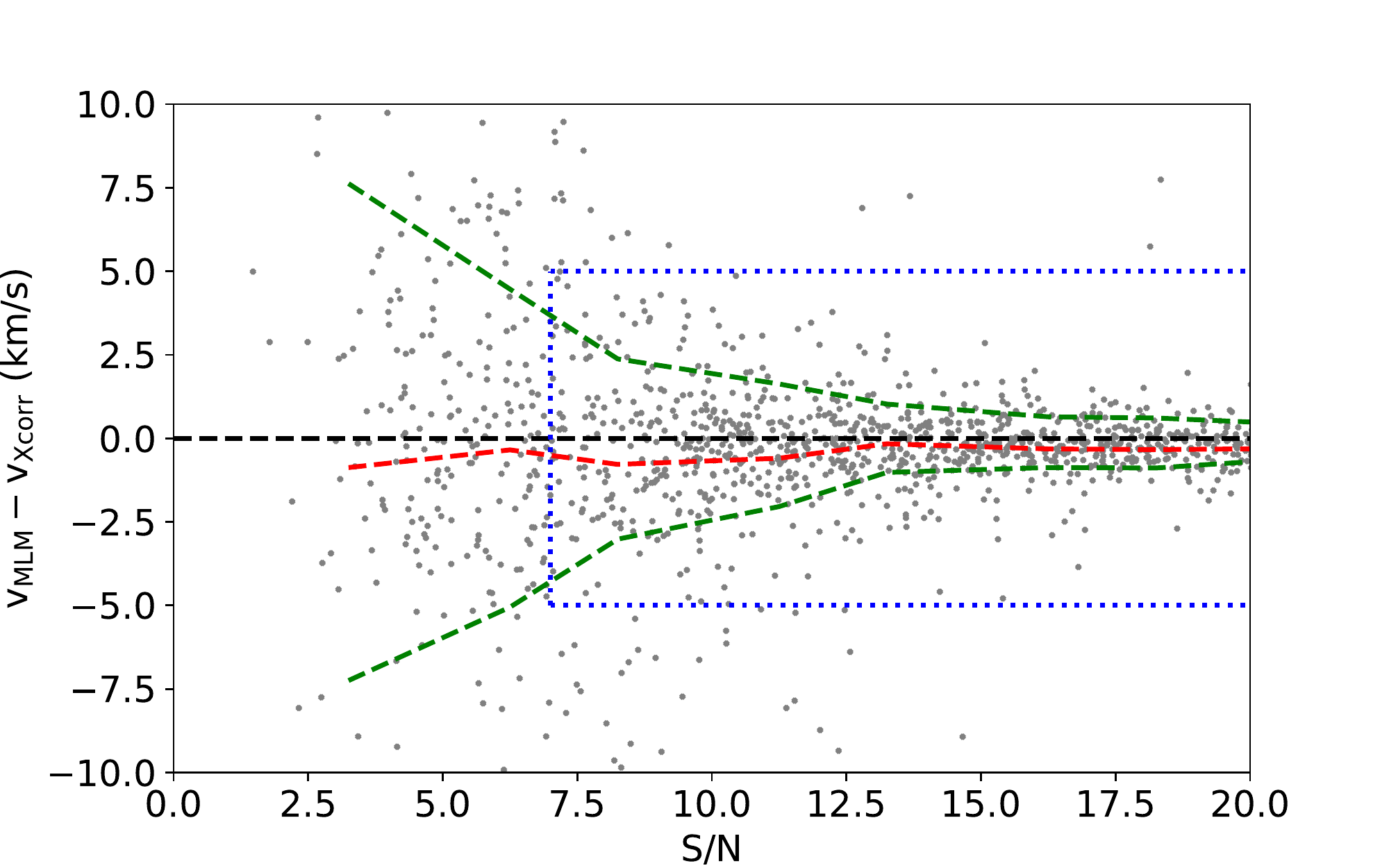}
\caption{{ Differences in \vlos\, at $\rm S/N<20$, as measured with the maximum likelihood (v$_\mathrm{MLM}$) and the cross-correlation (v$_\mathrm{Xcorr}$) methods on the same individual spectra, selected to have \vlos\ in the range expected for Sculptor stars ($70-150$\,km/s). Dashed green lines are the mean values above and below zero (dashed black line), and the dashed red line is the mean of all velocity differences. The vertical dotted blue line shows the minimum acceptable $\rm S/N=7$, and the horizontal dotted blue lines show the limits for outliers, at $\pm5$\,km/s.}}
\label{sncut}
\end{figure}

\begin{figure}
\centering
\includegraphics[width=1\linewidth]{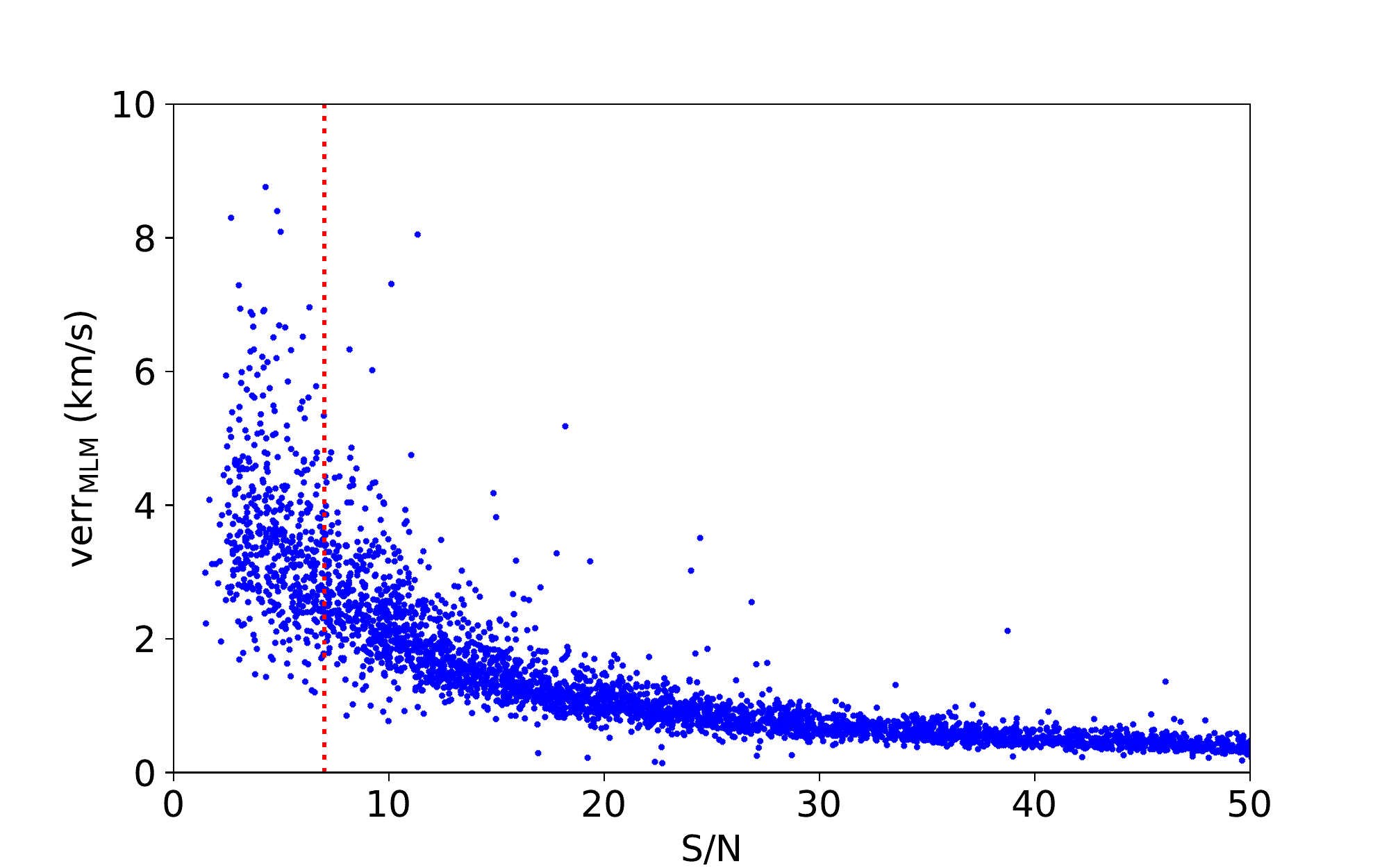}
\caption{ MLM velocity errors as a function of S/N for the full sample of individual VLT/FLAMES LR8 spectra, selected to have \vlos\ in the expected range for Sculptor stars ($70-150$\,km/s). The dashed red line indicates where the $\rm S/N>7$ cut, which is applied to the final selection, falls.}
\label{velerr}
\end{figure}

Two different methods of velocity measurements, MLM and cross-correlation, are compared in Fig.~\ref{sncut}, for individual velocity measurements in the likely \vlos\ membership range $70$--$150$~km/s.  The intrinsic scatter is expected to be $\lesssim1$~km/s, depending upon the S/N (see Fig.~\ref{fig-velerr}). It can be seen that although there is scatter, the majority of measurements lie in the expected range, and the scatter increases as the S/N decreases.  In Fig.~\ref{velerr} we can see how the MLM errors increased as the S/N decreased  with a similar spread to that seen in Fig.~\ref{sncut}, in both cases the scatter appears to expand significantly for S/N~$\lesssim~7$. Both the difference in MLM and Xcorr velocities and the MLM errors on velocity measurements were used as cuts for reliable \vlos\ measurements. The small number of stars at all S/N where the two velocity measurements differ well beyond their errors  (see Fig.~\ref{sncut}) were usually problems with the spectrum, for example, a poorly subtracted cosmic ray or sky line. This was quite a rare occurrence ($\sim$200 measurements out of  4646) and thus it was decided to neglect those spectra with a large difference ($>5$~km/s, which is $\sim3\sigma$) between the different velocity determinations. On some rare occasions a distant background object like a compact galaxy or an active galactic nucleus (AGN) found its way into the sample, these were sometimes also identifiable in this way. Figure~\ref{sncut} suggests that the measurements remain reliable until $\mathrm{S/N}\approx 7$ (vertical blue dotted line), even if the uncertainties are increasing. This was also consistent with what is seen in the MLM error distribution against S/N in Fig.~\ref{velerr}. After adopting the cuts $\rm S/N> 7$ and $|\rm v_\mathrm{MLM}-v_\mathrm{Xcorr}|<5$\,km/s, for the individual spectra, there were a total of 3312 VLT/FLAMES spectra for 1701 individual stars with reliable \vlos, within the likely membership range of the Sculptor dSph galaxy ($70-150$~km/s).

The \vlos\ measurements for two or more individual spectra with S/N$>7$ for the same star were combined, and weighted by $1/v_\mathrm{err}^2$. The mean measurements and their uncertainties are presented for stars with multiple measurements, as the mean and standard deviation of this combination. The cases of only a single measurement have a zero in the variance column. There are a significant number of stars in the whole VLT/FLAMES LR8 sample of 4646 spectra where the spectroscopic \vlos\ are clearly inconsistent with membership in the Sculptor dSph, for example  v$_{los} \gg 150$ km/s or  v$_{los} \ll 70$ km/s. Before \gaia\, only the position of the RGB in Sculptor could be used to select targets for spectroscopic follow-up, resulting in quite large numbers of non-members, especially in the outer regions. These non-members and their basic properties are recorded in Table E.2 (online only) in Appendix~E.

\begin{figure}
    \includegraphics[width=\linewidth]{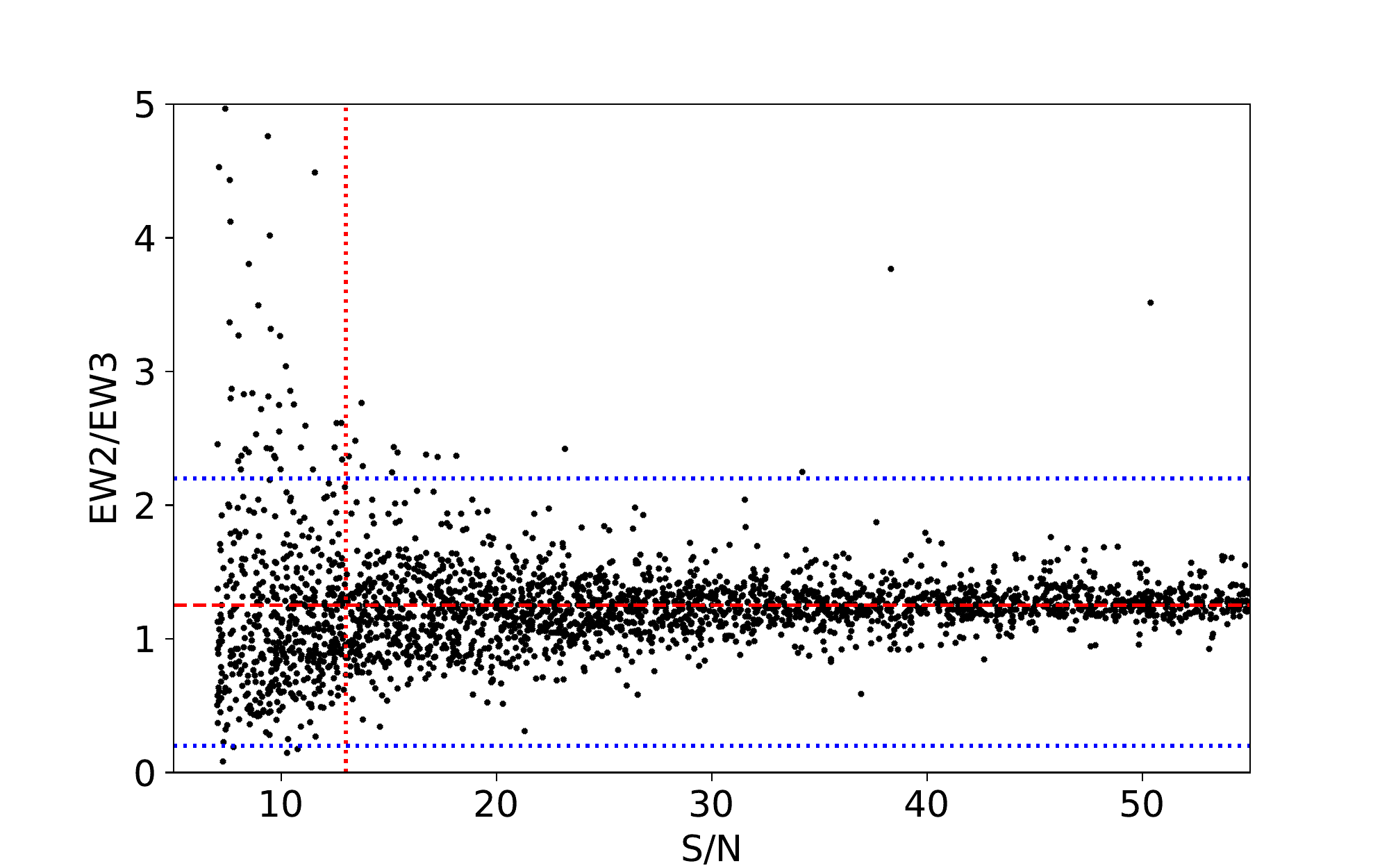}
    \caption{ Ratios of the two strongest CaT lines (EW2/EW3) as a function of S/N. The vertical dotted red   line is at S/N$=13,$ and dotted blue lines are limits of the reasonable values of the ratio at $\rm 0.2<EW2/EW3<2.2$. Outside this range there is most likely a problem with a poorly subtracted sky line interfering with the measurement of EW2 and/or EW3.}
    \label{fig:ew2}
\end{figure}

\subsubsection{The spectroscopic equivalent widths} 

Assuming that the target is an RGB member of the Sculptor dSph, then the equivalent widths (EWs) of the strongest two CaT lines, EW2 and EW3 at $\lambda\lambda$8542\,\AA\ \& $\lambda\lambda$8662\,\AA\ (see Fig.~\ref{specfeh}), can be used to determine \feh, through the well-established formalism (e.g. \citealt{Starkenburg10}, and references therein). The EWs of all three lines, EW1, EW2, and EW3, were measured using two methods: (1) integrating over the CaT lines; and (2) a more sophisticated fitting of both lines, assuming Gaussian profiles. In fitting the EWs the code computes a linear scale factor that best maps the Gaussian EWs onto the direct integration EWs and then applies that scale factor to the Gaussian EWs.  This effectively corrected for the difference between the real data where the CaT lines have (non-Gaussian) wings that depend on metallicity and a Gaussian model \citep{Battaglia08a}. The two approaches gave comparable results, but the fitting method provided more robust measurements, so this is what is presented here. All three lines were used for quality analysis of the spectrum, as peculiar ratios (e.g. EW1/EW2 and EW2/EW3) provide evidence for sky subtraction problems (see Fig.~\ref{fig:ew2}).

We looked through spectra at low S/N and also at cases with unusual \feh\ values, and determined that the S/N at which reliable metallicities can be obtained was somewhat higher than is required to measure reliable \vlos. Hence we adopted more severe cuts to our sample of spectra, $\rm S/N>13$, to obtain reliable \feh\ measurements. The measured EW ratios of different CaT lines were compared to their expected ratios, to ensure that individual lines were being measured accurately. This check flagged stars, mostly with low S/N, whose  measurements had intrinsically large uncertainties often with clear sky subtraction residuals (see the lowest S/N stars in Fig.~\ref{specsn}). In addition, there were cases where an unfortunately located cosmic ray led to problems with the sky subtraction, and/or an incorrect identification of the CaT lines. Thus, stars with strongly divergent EW1/EW2 or EW2/EW3 (see Fig.~\ref{fig:ew2}) ratios were removed, as were stars with unusually large EW2 or EW3 values, well beyond those expected. We were careful to ensure that the limits did not remove any realistic but unexpected measurements. This gave us a sample of 1414 stars with reliable \feh\ measurements, assuming the stars are members of Sculptor dSph, which will be determined in Sect.~\ref{sec:mem}.

\begin{figure*}
    \includegraphics[width=\linewidth]{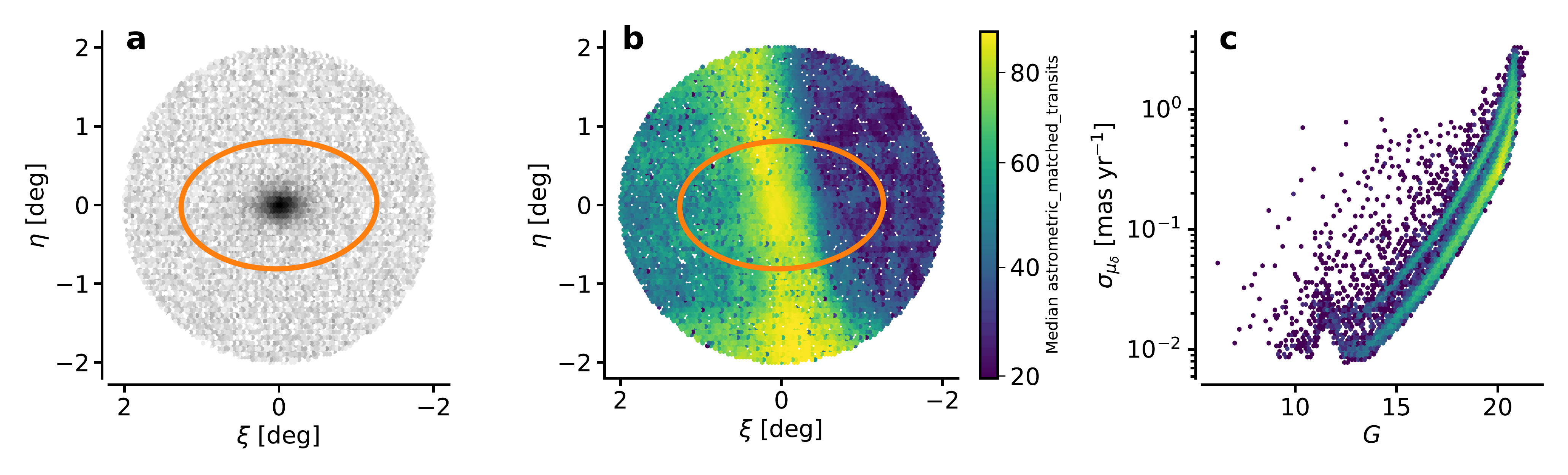}
    \caption{\gdr{3}\ data within two degrees of the centre of Sculptor.  (a)~Distribution of sources on the sky. (b)~Median number of astrometric observations (\texttt{astrometric\_matched\_transits}) matched to a given source. (c)~Uncertainty on \pmdec\ as a function of $G$. The sequence at higher values corresponds to the area in panel (b) with fewer visits. The orange ellipse in panels (a) and (b) indicates the nominal tidal radius of Sculptor.}
    \label{fig:sclgaiascans}
\end{figure*}

As with the velocities (Sect.~\ref{sec:specvel}), we combined the multiple measurements of \feh\ determined from EWs from independent spectra for the same stars, where the mean is weighted by the S/N. The errors on the \feh\ came from assuming that the intrinsic error on the sum of the two strongest lines (EW2+EW3) is 6/(S/N), where the constant is determined from comparing repeat measurements, and then this error was put into the metallicity equation with these limits added and subtracted for each measurement to obtain the error on the \feh\ determination \citep{Battaglia08a, Starkenburg10}. The mean measurements and their uncertainties are presented as a single value for stars with multiple measurements, as the mean and standard deviation of this combination. The cases of only a single measurement have zero in the variance column.

A sub-sample of our LR8 measurements also had VLT VLT/FLAMES High Resolution (HR, R$\sim$20 000) abundance analysis \citep{Hill19}. These HR observations allowed direct measurements of numerous Fe~I and Fe~II lines and thus provided a good check of our metallicity calibration (see Appendix~\ref{app:uves}). In addition, they provided abundances for a number of other elements, including Ca. From this comparison we see a possible,  small, offset in the measurements of the two different works, such that the \feh\ from VLT/FLAMES LR8 are $\sim 0.1$~dex more metal-rich than that measured by VLT/FLAMES HR10 (see Fig.~\ref{uves}). There is no obvious bias at play, there are no trends in the offset with metallicity, [Ca/H], or the G magnitudes of the stars (see Fig.~\ref{uvescat}). The [Ca/H] has a well defined trend with \feh\ in Sculptor \citep[e.g.][]{Hill19, Shetrone03}, so if this were the issue we would expect a different slope in Fig.~\ref{uves}a.
The scatter is within the uncertainties expected from the errors, and for two different methods, differently calibrated, such minor systematic offsets are not unexpected.

\subsubsection{Other surveys}

There have been several other major low/medium-resolution spectroscopic surveys of the Sculptor dSph made with other telescopes. For example, the Michigan/MIKE Fiber Spectrograph (MMFS) at the Magellan observatory was used to make a survey of 1365 likely members observed in the Mg-triplet spectral range \citep{Walker09}, and 262 of these have not been observed by VLT/FLAMES. In addition, there has been a Keck DEep Imaging Multi-Object Spectrograph (DEIMOS) study of 388 stars over a wide wavelength range using all the information in the faint lines \citep{Kirby09}. There are nine stars in the Keck sample that are neither in the Walker et al. sample nor in our VLT/FLAMES sample. These missing stars are typically fainter stars in the central region of the Sculptor dSph. Undoubtedly, new more complete samples will become available in the near future when 4MOST starts operation (e.g.~the 4DWARFS survey, PI: \'A. Sk\'ulad\'ottir, and the 4MOST low-resolution halo survey, PIs: E.~Starkenburg \& C.~Worley). The Subaru Prime Focus Spectrograph survey will also include Sculptor. The work presented in this paper should be useful for future surveys, using the overlapping observations to test calibration, and serve as a baseline for future galaxy-wide studies of the properties of variability and binary systems in Sculptor.

\begin{figure*}
    \centering
    \includegraphics[scale=0.4]{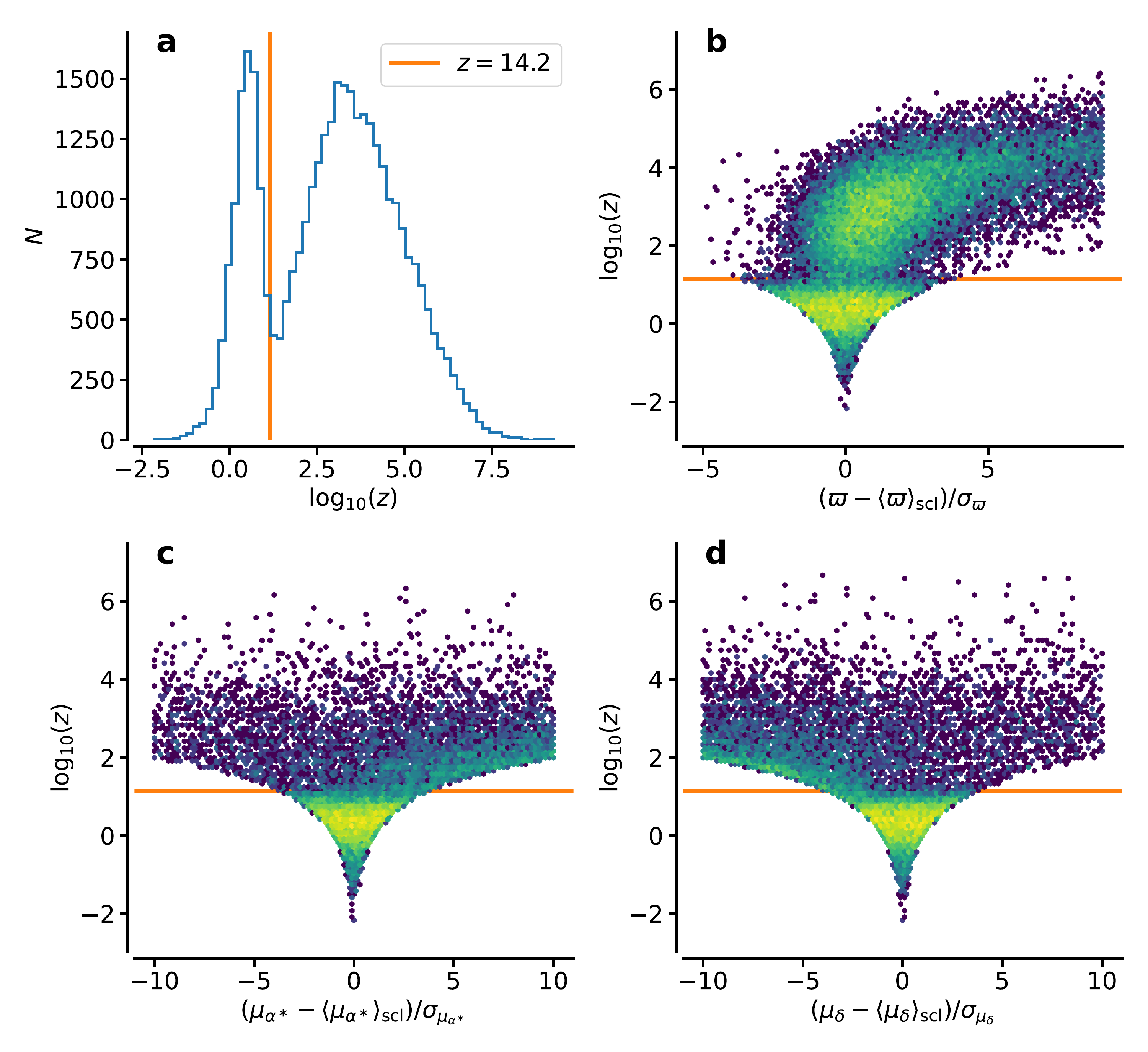}
    \caption{Sculptor membership score, $z$. (a)~Histogram of $\log_{10}(z)$ for all sources in the Sculptor field, where Sculptor members are expected to have $z<z_\mathrm{lim}=14.2$. The Distribution of $\log_{10}(z)$ is shown as a function of: (b)~$(\varpi-\langle\varpi\rangle_\mathrm{scl})/\sigma_\varpi$; (c) $(\pmra-\langle\pmra\rangle_\mathrm{scl})/\sigma_{\pmra}$; and (d) $(\pmdec-\langle\pmdec\rangle_\mathrm{scl})/\sigma_{\pmdec}$. The orange line always indicates $z_\mathrm{lim}=14.2$, which is the equivalent of $3\sigma$ for a 1D normal distribution, or the 99.9\% tile equivalent of the theoretical distribution of Sculptor members.}
    \label{fig:zscore}
\end{figure*}

\section{Membership selection}\label{sec:mem}

The \gdr{3} catalogue provides uniquely accurate information on the motions of stars in the plane of the sky. Even at the distance of the Sculptor dSph, the \gaia\ astrometry makes it possible to distinguish Galactic stars from Sculptor members. For individual stars we have parallax measurements with detailed error analysis; exquisitely accurate photometry, which gives the position of a star in a CMD; proper motions in right ascension (RA) and declination (Dec.) and their errors. The combination of all this information is a powerful discriminator for membership of a star in the Sculptor dSph. Not invoking too much prior information allows an unbiased assessment to be made of where the limits of membership lie. This is especially important in the outer regions of Sculptor. 

We describe here how stars belonging to the Sculptor population were selected on the basis of their \gdr{3} proper motions and parallaxes, and the los velocities measured with VLT/FLAMES. { The membership selection code is available on GitHub\footnote{\tiny \url{https://github.com/agabrown/sculptor-dwarf-galaxy-gaiadr3}}.}

\subsection{\gdr{3} data in the Sculptor field}

All \gdr{3} sources in the Sculptor field that have parallaxes, proper motions, and \bpminrp\ colours were selected according to the query listed in Appendix \ref{app:query}. Figure~\ref{fig:sclgaiascans} shows the distribution of these sources on the sky and the median number of astrometric measurements per source (\texttt{astrometric\_matched\_transits}). Figure~\ref{fig:sclgaiascans}b shows that Sculptor is located in a region of the sky with a rather uneven distribution in the number of observations collected per source. The core of Sculptor is very well observed, but the area west of Sculptor is less well covered. This leads to the difference in proper motion errors  illustrated Fig.~\ref{fig:sclgaiascans}c, which shows the uncertainty on the proper motion in Dec. versus $G$-band magnitude. The sequence at larger values of the uncertainties corresponds to the less well observed sky area west of Sculptor. A similar effect is seen in the RA proper motion uncertainties. The parallax uncertainties show no such split. However, this dichotomy in the uncertainties has no negative effect on the selection of Sculptor members.

A first exploratory analysis of the sources in the direction of Sculptor dSph was carried out by examining their distribution on the sky, in parallax, in the CMD and in the normalised corrected flux excess factor $|C^*/\sigma_{C^*}|$, as defined by \cite{Riello21}. High values of $|C^*/\sigma_{C^*}|$ can indicate crowding issues or the presence of (extended) non-stellar sources. A large fraction of sources at high values of $|C^*/\sigma_{C^*}|$ were found to be listed as quasi-stellar objects (QSOs) or galaxy candidates in the \gdr{3} catalogue, thus indicating that source crowding is not a major concern in the Sculptor field. This is confirmed by the value of the $M_{10}$ indicator \citep{Cantat22}, which is $21.2$ for the Sculptor field, indicating a high completeness of the \gdr{3} catalogue out to the \gaia\ survey limit at $G\sim21$ and a corresponding lack of crowding issues.

\begin{figure*}
    \centering
    \includegraphics[width=\linewidth]{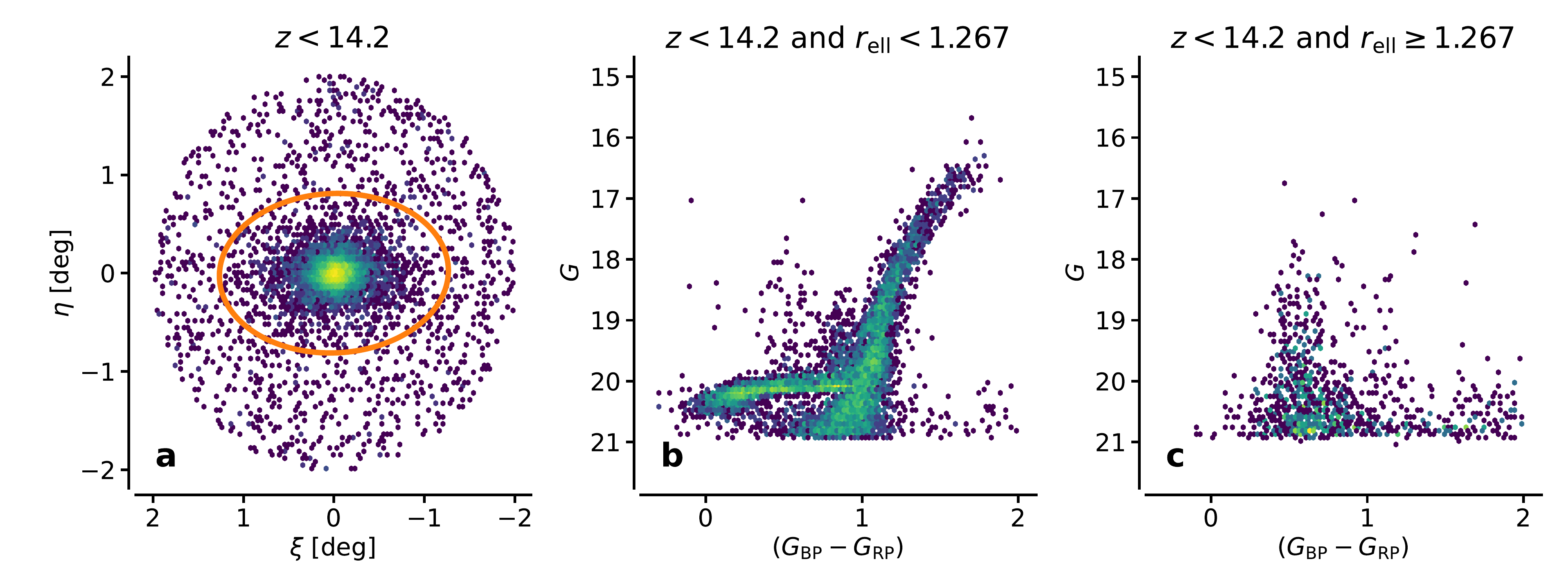}
    \caption{Sculptor members selected on the basis of their parallaxes and proper motions. (a) Distribution on the sky. The orange ellipse indicates the nominal tidal radius of Sculptor. (b) and (c) CMDs for the Sculptor members located inside and outside the nominal tidal radius ($r_\mathrm{ell}=1.267^\circ$), respectively.}
    \label{fig:3dmemprops}
\end{figure*}

A preliminary selection of Sculptor stars was made on the basis of their proper motions and parallaxes, with the latter corrected for the parallax bias using the recipe from \cite{Lindegren2021}. The results show that the mean parallax bias correction applied to the selected Sculptor stars was $\sim17$\,\muas\, corresponding to the mean parallax bias for QSOs in \gdr{3} \citep{Lindegren2021}. However, the mean parallax of the \gaia\ celestial reference frame objects (QSOs and AGNs) from \gdr{3} \citep{Klioner2022} in the Sculptor field is $-39$\,\muas, suggesting that a larger correction should be applied. In addition, the raw parallaxes of the preliminary list of members have a median value of $13.6$\,\muas, very close to the nominal Sculptor parallax $11.9\pm0.2$~\muas, based on the distance modules listed in \citet{Battaglia22}\footnote{Applying either the $17$ or $39$\,\muas\ parallax correction would result in a distance modulus for Sculptor of $17.6$ or $16.4,$ which are both clearly incompatible with the observed apparent brightness of the Sculptor horizontal branch at $G\approx20.1$.}. We therefore decided to work with the raw parallaxes and ignore the bias corrections.

To select Sculptor member stars, we assumed the proper motion and the parallax corresponding to the distance listed in \cite[see our Table \ref{tablepm}]{Battaglia22} and calculated a membership score, $z_i$, for each \gdr{3}\ source, $i$, in the Sculptor field: 
\begin{equation}
    z_i = \mathbf{v}_i^\prime (\mathbf{C}_i+\mathbf{D})^{-1} \mathbf{v}_i\,,
    \label{eq:zscore}
\end{equation}
where $^\prime$ indicates the transpose of the vector, $\mathbf{v}_i$:
\begin{equation}
    \mathbf{v}_i = \begin{pmatrix}
        \varpi_i - \langle\varpi\rangle_\mathrm{scl} \\ 
        \mu_{\alpha*,i} - \langle\mu_{\alpha*}\rangle_\mathrm{scl} \\
        \mu_{\delta,i} - \langle\mu_\delta\rangle_\mathrm{scl}
    \end{pmatrix}
.\end{equation}
The matrix $\mathbf{C}_i$ is the covariance matrix for the parallax and proper motions of source $i$ as listed in \gdr{3}. The matrix $\mathbf{D}$ contains the squares of the uncertainties on the mean parallax and proper motions of Sculptor on its diagonal and all other elements are zero.

The membership score $z$ should be a $\chi^2$ distribution with three degrees of freedom for Sculptor members, and an upper limit on $z$ can be used to select members. 
Figure~\ref{fig:zscore}a  shows a histogram of $z_i$ that reveals a clear presence of two distinct groups of sources, at low (Sculptor members) and high (non-members) values of $z$; the orange line shows the value $z_\mathrm{lim}=14.2$, which is the limit for a $\chi^2_3$ distribution corresponding to a $3\sigma$ limit for a 1D normal distribution. Figure~\ref{fig:zscore}b shows that for $z<z_\mathrm{lim}$ the source parallaxes are normally distributed around the mean Sculptor parallax, while above the limit the distribution strongly deviates from normal. This can be explained as a population of sources at effectively zero parallax concentrated around the proper motion of Sculptor ($z<z_\mathrm{lim}$) and a population of Milky Way stars with different proper motions (leading to high values of $z>z_\mathrm{lim}$) and significant positive parallaxes (stars at close distances). The sharp boundary suggests that foreground stars are effectively removed without introducing an explicit limit on the parallax $\varpi$ or $\varpi/\sigma_\varpi$. Figs.~\ref{fig:zscore}c, d show  that in proper motions the boundary at $z_\mathrm{lim}=14.2$ is not as well defined, with a smooth transition of large positive \pmra\ values and large negative \pmdec\ values. This indicates the presence of interlopers belonging to the Milky Way stellar population.

\subsection{Sculptor membership selection}
Selecting Sculptor stars simply according to $z<14.2$ results in $8861$ members. We removed sources with $|C^*/\sigma_{C^*}|>7$ and values of the re-normalised unit weight error (RUWE) parameter above $1.25$. These high RUWE sources stand out as having much larger parallax errors than the mean for sources at similar brightness. These selections leave $8375$ sources as Sculptor members. The distributions of the parallaxes and proper motions are consistent with 1D normal distributions, with a slight excess of negative $\mu_\delta$ values (see Fig. \ref{fig:zscore}b). The widths of the distributions suggest that the uncertainties on the parallaxes and proper motions in RA and Dec. are underestimated by $7$, $12$, and $14$ percent, respectively, consistent with other findings in the literature (e.g.\ Table 1 in \citealt{Fabricius2021}). The median parallax is $0.0136$~mas. The weighted mean proper motions are $\pmra=0.095\pm0.002$ and $\pmdec=-0.154\pm0.002$~\masyr, with a correlation coefficient of $-0.36$ (like the parallaxes, the proper motions are not corrected for systematic errors). The distributions of the selected sources on the sky and in the CMD are shown in Fig.~\ref{fig:3dmemprops}. The CMD  in Fig.~\ref{fig:3dmemprops}c shows the stars outside the nominal tidal radius of Sculptor. { This shows that our adopted definition of the tidal radius \citep{Irwin95} is consistent with the \gaia\ selection as the outside population looks entirely consistent with foreground contamination.}
The distribution is representative of Milky Way stars in the direction of Sculptor as is confirmed by selecting stars on the same parallax and proper motion criteria in a control field 4 degrees to the east of Sculptor.

\begin{figure*}
    \centering
    \includegraphics[scale=0.45]{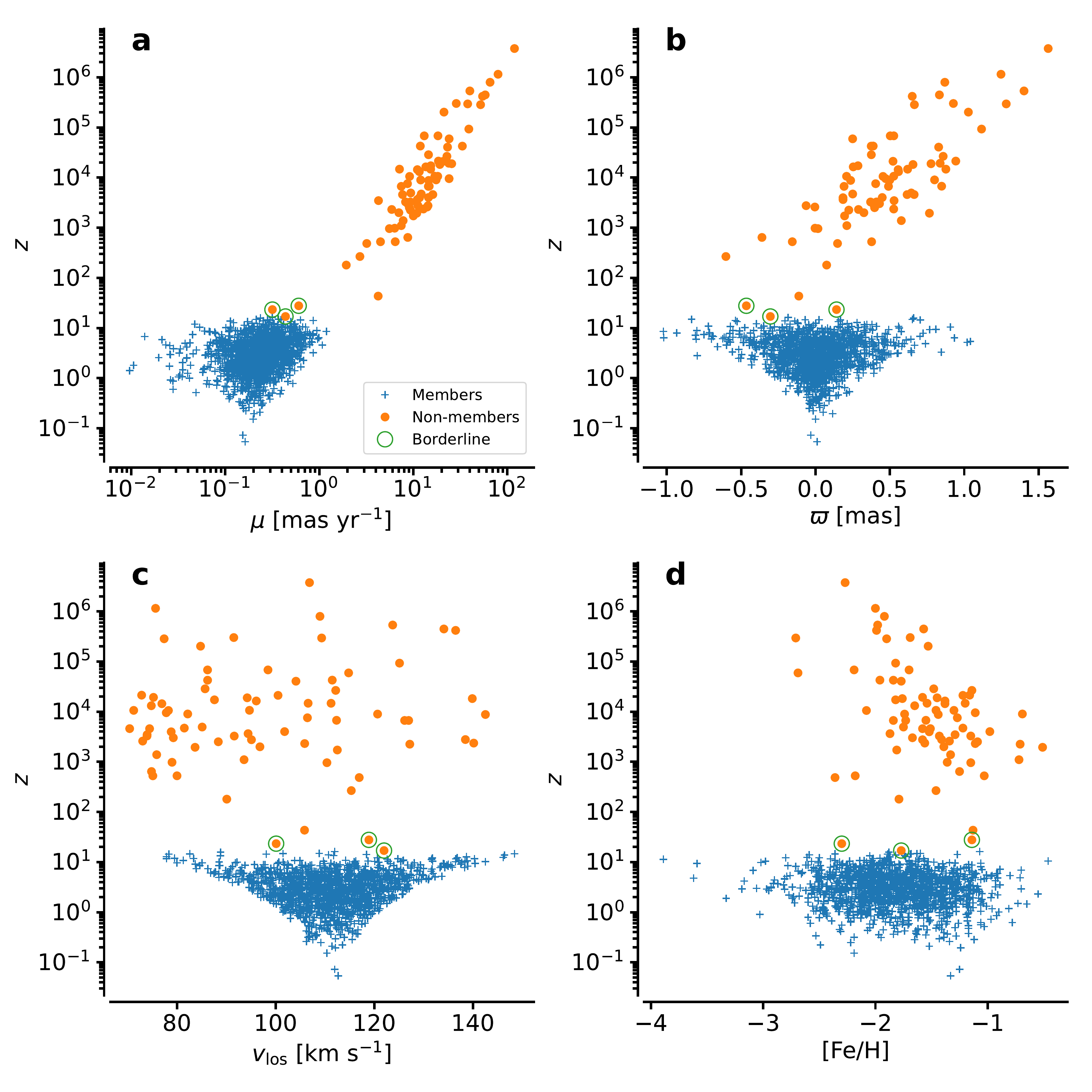}
    \caption{ $z$ membership score of stars in the VLT/FLAMES LR8 sample as a function of: (a)~total proper motion, $\mu$; (b)~parallax; (c)~los velocity; and (d)~metallicity{ (using only the 1339 reliable measurements)}. This shows the clean separation between Sculptor members and non-members. The three orange dots with green circles are borderline cases. }
    \label{fig:flamesmemprops}
\end{figure*}

\begin{figure*}
    \centering
    \includegraphics[scale=0.4]{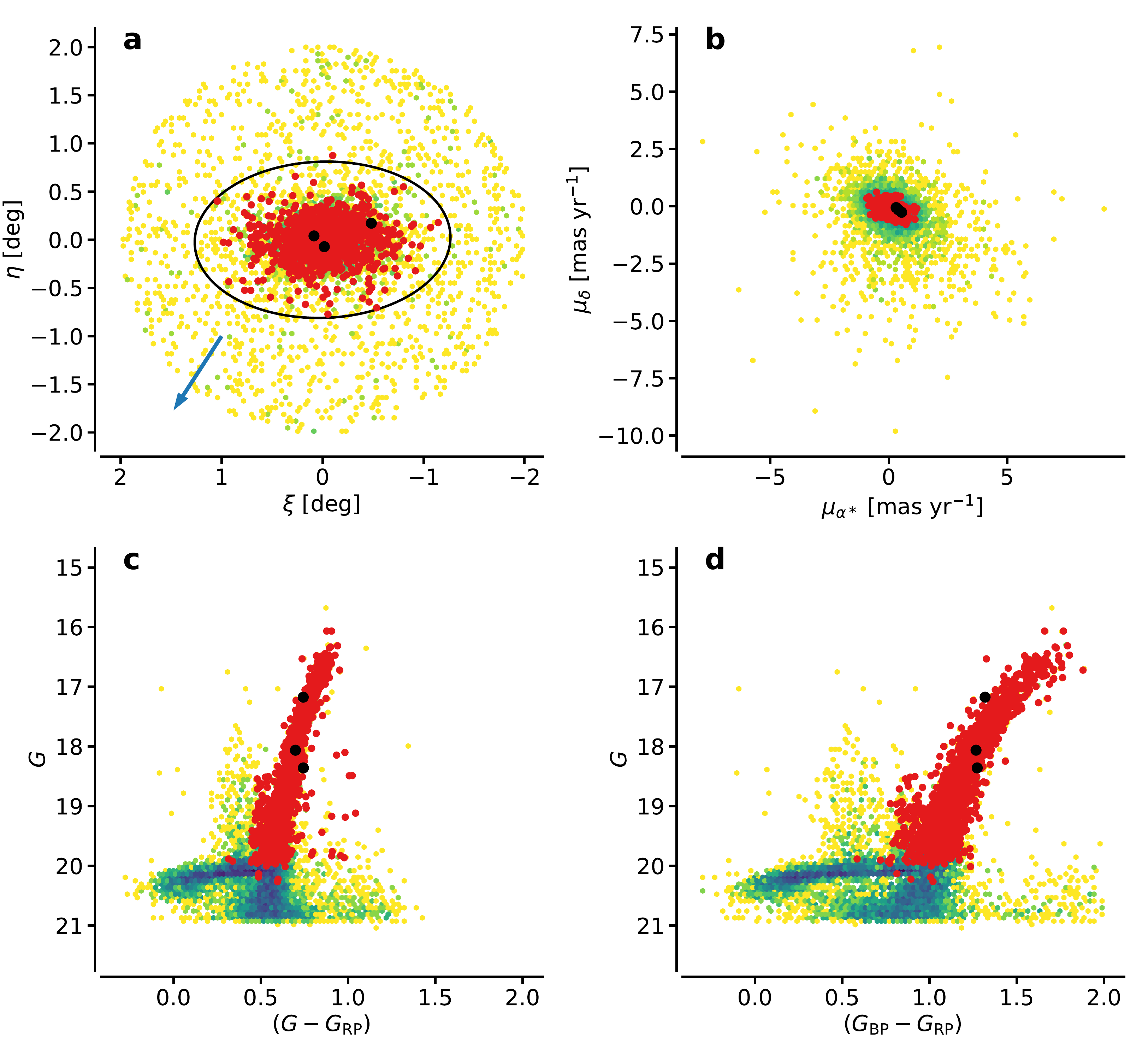}
    \caption{Sculptor members in the VLT/FLAMES LR8 sample (red), as well as all \gdr{3}\ Sculptor members selected on the basis of parallax and proper motion only (yellow-green colour coding): (a) on the sky, where the black ellipse indicates the nominal tidal radius and the blue arrow the direction of proper motion of Sculptor in the plane of the sky; (b) in proper motion; and (c) and (d) in the CMDs. The larger black symbols indicate the borderline cases between non-members and members identified in Fig.~\ref{fig:flamesmemprops}.}
    \label{fig:flamesmemskycmd}
\end{figure*}

Next we selected Sculptor member stars from the VLT/FLAMES LR8 survey by also accounting for the los velocity, \vlos. We used the above derived uncertainty inflation factors ($1.07$, $1.12$, and $1.14$ for parallax and proper motion in RA and Dec., respectively) and the weighted mean proper motions of Sculptor and their covariance matrix. We calculated the $z$ value again according to Eq.~\ref{eq:zscore} but with
\begin{equation}
    \mathbf{v}_i = \begin{pmatrix}
        \varpi_i - \langle\varpi\rangle_\mathrm{scl} \\ 
        \mu_{\alpha*,i} - \langle\mu_{\alpha*}\rangle_\mathrm{scl} \\
        \mu_{\delta,i} - \langle\mu_\delta\rangle_\mathrm{scl} \\
        v_{\mathrm{los},i} - \langle v_\mathrm{los}\rangle_\mathrm{scl}
    \end{pmatrix}
\end{equation}
and
\begin{equation}
    \mathbf{D} = \begin{pmatrix}
        \sigma_{\varpi,\mathrm{scl}}^2 & 0 & 0 \\
        0 & \mathbf{C}_{\mu,\mathrm{scl}} & 0 \\
        0 & 0 & \sigma_{v_\mathrm{los,scl}}^2 + \sigma_\mathrm{scl}^2
    \end{pmatrix}
,\end{equation}
where $\mathbf{C}_{\mu,\mathrm{scl}}$ is the covariance matrix of the mean proper motions for Sculptor (derived above) and $\sigma_\mathrm{scl}$ is 
the los velocity dispersion of Sculptor. We use $\langle v_\mathrm{los}\rangle_\mathrm{scl}=110.6\pm0.5$\,\kms\ and $\sigma_\mathrm{scl}=10.10$~\kms \citep{Battaglia08b}. {The covariance matrices $\mathbf{C}_i$ include an additional diagonal term, corresponding to the uncertainty on the measured radial velocity for star $i$.} The upper limit on $z$, $z_\mathrm{lim}=16.3$, now follows from the $\chi^2_4$ distribution. The membership selection results in { 1604} out of 1701 stars selected as LR8 radial velocity members of Sculptor { with an additional three stars as borderline cases}. Thus, out of the Sculptor VLT/FLAMES LR8 spectroscopic \vlos\ members (with velocities in the range 70--150\,km/s), 94\% are confirmed to be member stars when \gdr{3} information is taken into account. There were no stars with velocities outside this range in our spectroscopic sample that had \gdr{3} properties consistent with membership in Sculptor dSph. On the other hand, out of the 1609 stars  in our spectroscopic sample that are members according to \gdr{3} 1604 or 99.7\% are confirmed as Sculptor members once \vlos\ is also included.

Figure~\ref{fig:flamesmemprops} shows the $z$ membership score as a function of the total proper motion, parallax, los velocity, and metallicity. It is clear from this figure that most of the { 95} non-members { from the VLT/FLAMES spectroscopic sample with $\rm70\,km/s>\vlos<150\,km/s$ } are stars with large proper motions { (see Fig.~\ref{fig:flamesmemprops}a)}, clearly { not} belonging to the { Sculptor dSph}. The non-members do not stand out in los velocity { (Fig.~\ref{fig:flamesmemprops}c)} but do tend to be more metal-rich { (Fig.~\ref{fig:flamesmemprops}d)} and have larger parallaxes { (Fig.~\ref{fig:flamesmemprops}b)} than the bulk of the Sculptor members in the VLT/FLAMES sample. { It is the combination of all these different measurements that make the most powerful membership determination.} The three green circled orange symbols in all the panels of Fig.~\ref{fig:flamesmemprops} indicate stars that are formally non-members (according to the $z$-score but in proper motion, { arguably the most stringent criteria, combined with \vlos,} are consistent with being Sculptor members and well separated from the other non-members. These three stars could thus be considered borderline cases and may well be members of Sculptor. Their \gaia\ properties and their spectroscopic properties are given in the results table (Table E.1) in Appendix~E.

Figure~\ref{fig:flamesmemskycmd} shows the distribution of the VLT/FLAMES members on the sky, in proper motion, and over the CMD, where the background colour coding indicates the density of Sculptor sources selected on the basis of their proper motion and parallax only. In Fig.~\ref{fig:flamesmemskycmd} we also show where the borderline cases identified in Fig.~\ref{fig:flamesmemprops} lie. They also show properties consistent with membership.

Figures \ref{fig:flamesmemskycmd}a and 12b show how the members are concentrated in the central regions of the sky (by design in the VLT/FLAMES survey) and proper motion distributions for Sculptor. {It is clear from Fig.~\ref{fig:flamesmemskycmd}a that the nominal tidal radius used \citep[from][]{Irwin95} is consistent with the confirmed member distribution in Sculptor. However, the spectroscopic follow-up did not extend much beyond this limit as there were very few likely members along the Sculptor RGB. In the northern-most VLT/FLAMES field (Scl025) there were spectra of 4 stars that passed the quality criteria and had radial velocities consistent with membership in Sculptor dSph, but only 1 of these also had proper motions and parallax consistent with membership (scl\_25\_031). This single star is only just beyond the nominal tidal radius, as can be seen in Fig.~\ref{fig:flamesmemskycmd}a. Larger fields of view, like that provided by 4MOST, will be best suited to search for extra-tidal stars  (4DWARFS; Skúladóttir et al. {ESO Messenger}, in press).}

In Figs.~\ref{fig:flamesmemskycmd}c and d, the CMDs show the location of the VLT/FLAMES members on the RGB. In both CMD panels the sources bluewards of the  RGB at magnitudes between $G=20$ and $G=18.5$ are probably dominated by Milky Way interlopers (see Fig.~\ref{fig:3dmemprops}). However, these (possible) interlopers are indistinguishable from Sculptor members within our current criteria. In addition, it can be seen that there are $\sim$12 stars in Fig.~\ref{fig:flamesmemskycmd}c that lie to the red of the Sculptor RGB in $(G-\grp)$, but they appear on the RGB in $(\gbp-\grp$) in Fig.~\ref{fig:flamesmemskycmd}d. These are stars with a neighbour in \gdr{3} within 3~arcsec, which can lead to blending of the \gbp\ and \grp\ images of both sources and thus cross-talk between the integrated fluxes for these cases. This means that both \gbp\ and \grp\ are biased, with an overestimated flux due to the neighbouring source. This also shows up in $G-\gbp$ as too blue. In $\gbp-\grp$ the biases compensate. We see no effect in the spectra, so either the 1.2~arcsec VLT/FLAMES fibres  do not see the neighbouring object, or it is a sufficiently different source that does not affect our CaT spectra.

\subsection{Comparing with previous membership determinations}

Other groups have made membership selections for Sculptor based on \gdr{2} and \gdr{3} using different criteria \citep[e.g.][]{Simon18, McV20, Battaglia22}, determining the probability of membership based upon the expectations of a Sculptor-like population in this region of the sky. Typically these selections make assumptions, for example on the expected distribution in the sky and in the CMD that may potentially lead to small numbers of members being overlooked in sparsely populated regions. This is not critical for global studies to compare the properties of many different dwarf galaxies. Here, as we want to explore the Sculptor dwarf galaxy in detail, we preferred not to make any assumptions on either the spatial distribution of the stars nor their location in the CMD, so that we 
can look in detail at the selection and not rule out the presence of unusual members or possible members beyond the nominal tidal radius.
There are slight differences between our \gaia-only selection of { 8375} likely members and the 6576 sources that \citet{Battaglia22} assigned a probability of membership, P $>95\%$, but they are small enough not make any significant difference to the results.
Of the 1604 members in our \gdr{3}-LR8 sample, 32 (mostly faint) stars are not in the Battaglia et al. catalogue of members. Some of these can be explained by the application of different quality cuts on the \gaia\ sources (e.g. AGN flags, duplicates). They are all on the RGB in the CMD and within the nominal tidal radius. There are 5 stars that both Battaglia et al. and our selection find to be likely \gaia\ members but they are VLT/FLAMES \vlos non-members. These are therefore extremely small differences. Both methods are approaching the border between members and non-members in slightly different ways and so small inconsistencies are not surprising. 

\subsection{Completeness of the \gdr{3}-LR8 survey}

 The spectroscopic follow-up covers the full width (and beyond) of the RGB (see Fig.~\ref{fig:flamesmemskycmd}d), as it was designed to do. It also populates well the physical centre of Sculptor on the sky, where most of the stars are to be found (Fig.~\ref{fig:flamesmemskycmd}a). We have 1701 VLT/FLAMES LR8 spectra with \vlos\ broadly consistent with membership of the Sculptor dSph, $70$--$150$~km/s, and $\mathrm{S/N}>7$. Out of these, 1604 have \gdr{3} astrometry consistent with membership of the Sculptor dSph and 96 do not. In addition, there are three borderline stars that appear to be broadly consistent with Sculptor membership but formally do not make the cut in z. It is interesting to note that roughly 6\% of previously identified velocity members are actually non-members when the \gaia\ astrometry is included. In the outer regions there are very few stars and so each one is important in determining the properties of the sparsely populated regions, and here it is critical that members and non-members are correctly separated. 

\begin{figure}
\centering
\includegraphics[width=1\linewidth]{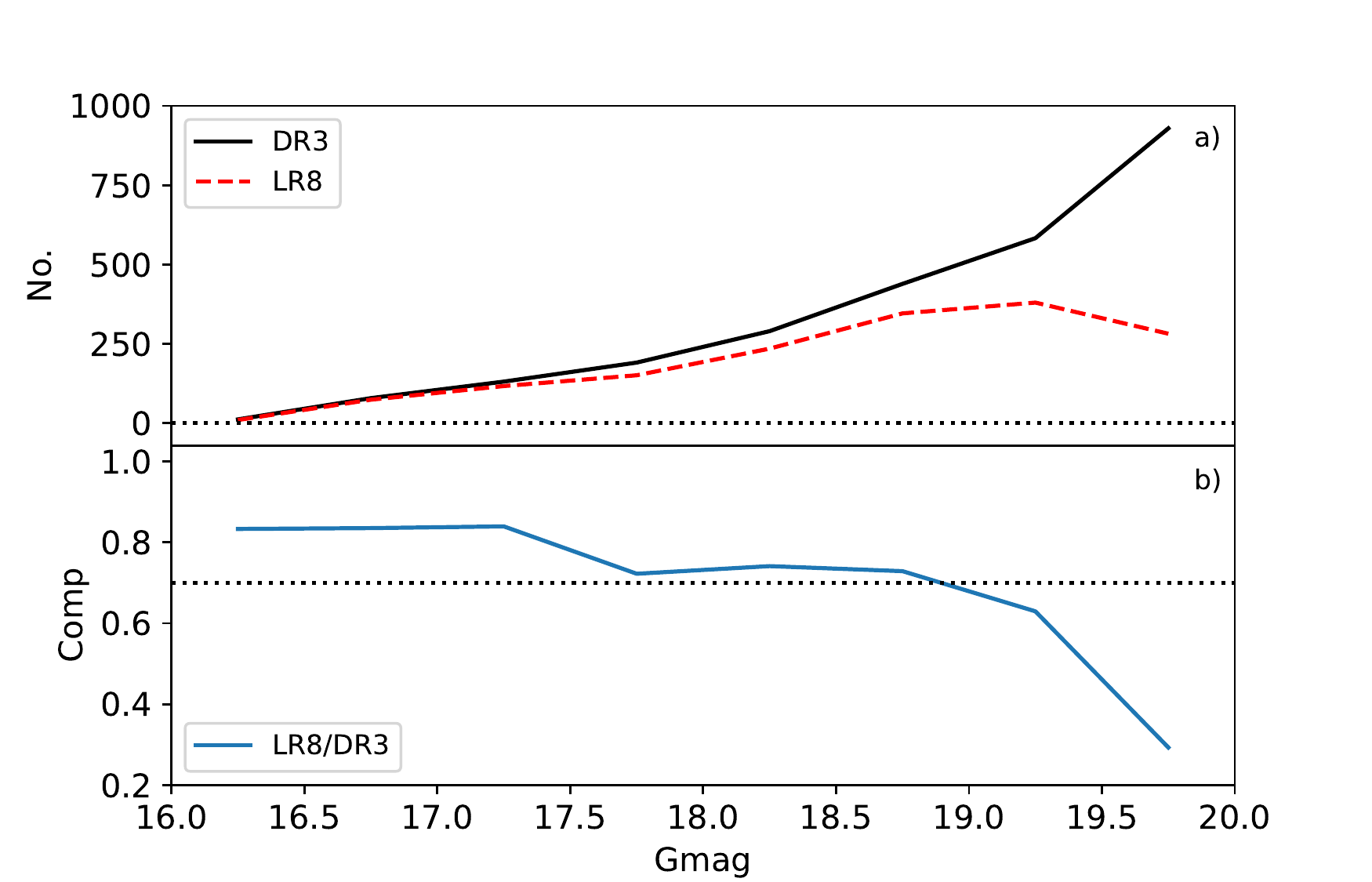}
\caption{Completeness of the VLT/FLAMES LR8 survey relative to \gdr{3} for Sculptor members as a function of \gaia\ G magnitude: (a)~the number of \gdr{3} astrometric members (solid black line) and those with VLT/FLAMES LR8 spectroscopic confirmation (dashed red line), with a black dotted line at null; (b)~the  fraction of \gdr{3} astrometric members that have a VLT/FLAMES LR8 spectrum with S/N$>${ 7} and \vlos\ consistent with membership in Sculptor. The black dotted line denotes 70\%.}
\label{figcomp}
\end{figure}

\begin{figure}
\centering
\includegraphics[width=1\linewidth]{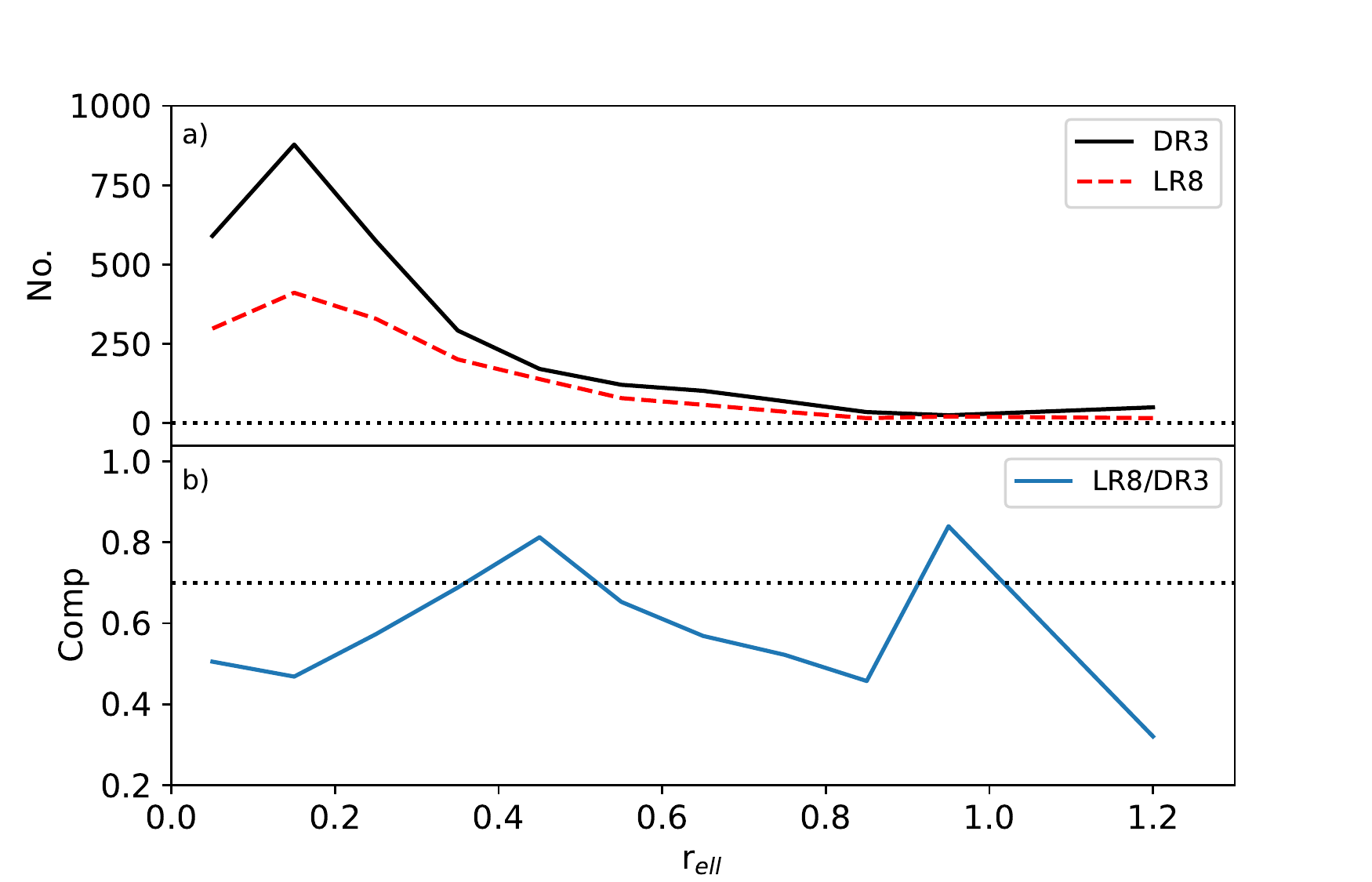}
\caption{Completeness of the VLT/FLAMES LR8 survey relative to \gdr{3} for Sculptor members as a function of elliptical radius (r$_{ell}$): (a)~the number of \gdr{3} astrometric members (solid black line), and those with VLT/FLAMES LR8 spectroscopic confirmation (dashed red line), with a black dotted line at null; (b)~ fraction of \gdr{3} astrometric members that have a VLT/FLAMES LR8 spectrum of sufficient S/N, $>${ 7}, and a \vlos\ consistent with membership. The dotted black line denotes 70\%.}
\label{figellcomp}
\end{figure}

In Fig.~\ref{figcomp} we plot an overview of the completeness as a function of \gaia\ G-magnitude of the \gdr{3}-LR8 sample. We observed $\sim55$\% of the \gaia\ selection with VLT/FLAMES in the range $G=16$--$20$~mag, and with $>70$\% completeness in the brightest range, $G<$18.75.  In Fig.~\ref{figellcomp} the same comparison is made as a function of elliptical radius. The completeness is quite uneven due to the decreasing numbers of stars going towards the outer sparse regions of Sculptor. The dip in the centre is due to the large numbers here and the difficulties in putting them all on to VLT/FLAMES fibres due to crowding. 
The completeness in the central, more metal-rich region of the galaxy is fairly similar in percentages to the outer, more metal-poor regions (see Fig.~\ref{figellcomp}); however, in total numbers, the central region lacks spectroscopic observations of more stars.

\begin{table}
\caption{Mean properties of the Sculptor dSph galaxy; the errors are the standard deviation.}             
\label{tablepm}      
\centering                                      
\begin{adjustbox}{width=\columnwidth,center}
\begin{tabular}{l c c}          
\hline\hline                        
Property & ~~Mean & ~~Previous\\    
\hline                                   
    $\rm \langle\feh \rangle$ & ~~{ -1.82} $\pm$ { 0.45} & ~~-1.58 $\pm$ 0.41$^{[1]}$ \\
    $  \rm \langle \vlos \rangle$ (km/s)& ~~{ 111.2} $\pm $ 0.25 & ~~110.6 $\pm$ 0.5 $^{[2]}$\\      
                              &                                 & ~~111.4 $\pm$ 0.1 $^{[3]}$\\  

     \vlos  (\feh$>-$1.7) & ~~{ 111.50} $\pm$ { 0.36}   & \\
          \vlos  (\feh$<-$1.7) & ~~{ 110.06} $\pm$ { 0.37} &     \\ 
        \vlos  (\feh$<-$2.2) & ~~{ 112.27} $\pm$ { 0.67} &     \\ 
        \vlos  (\feh$<-$2.5) & ~~{ 114.82} $\pm$ { 1.37} &     \\ 
\hline    

    $\langle\mu_\alpha\rangle_{sp}$ (mas/yr) & ~~{ 0.097}  $\pm$ { 0.006} &  ~~$0.099  \pm 0.002 ^{[4]}$ \\
    $\langle\mu_\delta\rangle_{sp}$  (mas/yr) & { -0.148} $\pm$ { 0.004}  &   $-0.159 \pm 0.002 ^{[4]}$  \\
    $\langle\varpi \rangle_{sp}$  (mas) & { 0.004} $\pm$ { 0.007} &   $- 0.013 \pm 0.004 ^{[5]}$ \\
\hline                                             
\end{tabular}
\end{adjustbox}
    \begin{tablenotes}
      \scriptsize
      \item $^{[1]}$ \citet{Kirby09}
      \item $^{[2]}$ \citet{Battaglia08a}
      \item $^{[3]}$ \citet{Walker09}
      \item $^{[4]}$ \citet{Battaglia22}, based on \gdr{3}
      \item $^{[5]
}$ \citet{Helmi18}, based on \gdr{2}
      
    \end{tablenotes}
\end{table}

\section{Results}

The Sculptor dSph is a very well-studied galaxy. Here we looked at a new combination of proper motions and spectroscopic velocities for 1604 individual RGB stars found over the full area of the galaxy on the sky, from the centre out to the nominal tidal radius (Fig.~\ref{fig:flamesmemskycmd}a). This combination provides a powerful method for accurately selecting Sculptor member stars, especially in the more challenging, sparse outer regions. This makes it useful to (re-)investigate the systemic velocity and other global properties on the basis of this new dataset (see Table~\ref{tablepm}). Furthermore, for a subset of 1339 Sculptor stars, whose spectra had S/N$>13$, we also could provide reliable metallicities, \feh. The values of \vlos\ and \feh\ for the whole spectroscopic dataset (including non-members) are given in Tables~E.1 and E.2 (online only) in Appendix~E. In addition, the \gdr{3} parallax and proper motions are also provided, as well the proper motion membership determination. We also include information about the non-member stars that originally appeared to be spectroscopic members. With our new, larger, and more accurate dataset, we combined the 3D velocities of individual stars with their measured metallicities and took a look at the chemo-dynamical properties of the stellar population in the Sculptor dSph.

\begin{figure}[ht]
\centering
\includegraphics[width=1.\linewidth]{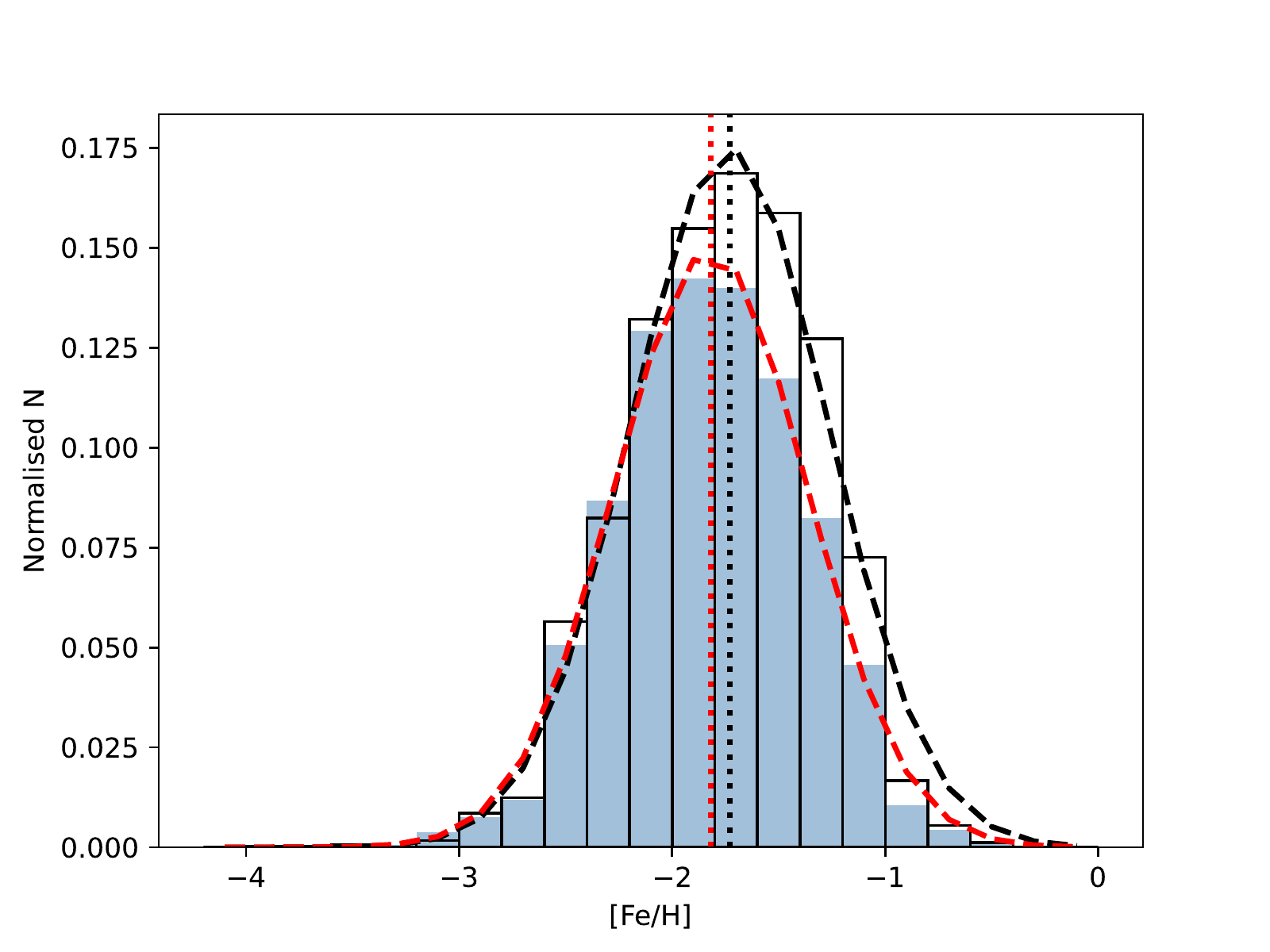}
\caption{ MDF for the 1339 metallicity measurements of RGB members of the Sculptor dSph (blue). The vertical dotted red line is the mean, $\rm \feh={ -1.82}$, as determined by a Gaussian fit to the distribution (dashed red line). The black-outlined histogram is corrected for the incompleteness at different elliptical radii, as given in
Fig.~\ref{figellcomp}. The mean metallicity shifts by about 0.1\,dex (dotted black line) to $\rm\feh=-1.73$, according to the Gaussian fit (dashed black line).}
\label{fehhist}
\end{figure}

\begin{figure*}[t]
\centering 
\includegraphics[width=1.\linewidth]{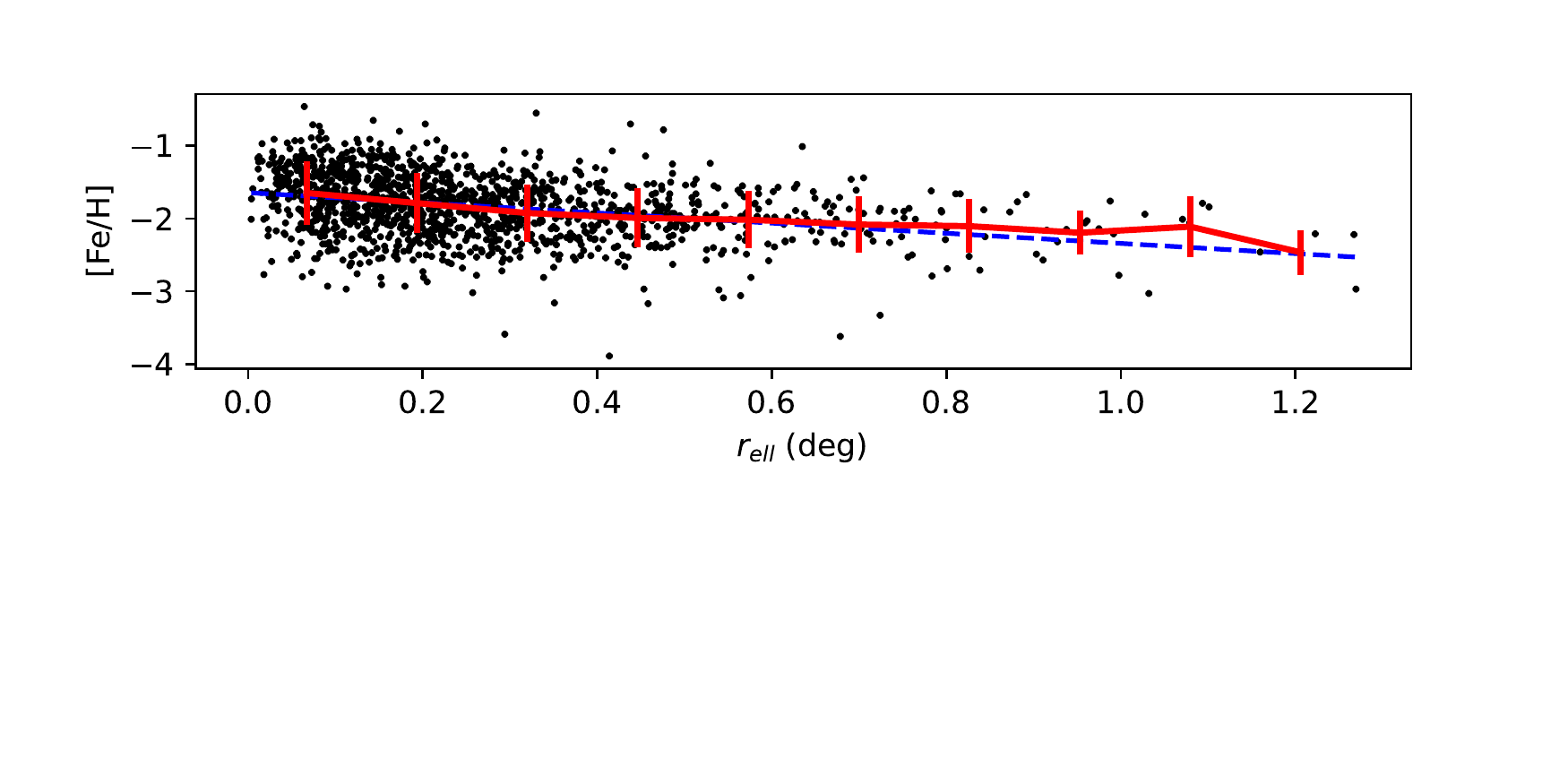}

\caption{ \feh\ with elliptical radius, r$_{ell}$, which is the projected semi-major axis radius, for the 1339 measurements in our Sculptor sample. The dashed blue  line shows the linear fit, which represents a gradient of $-0.7$dex~deg$^{-1}$. The red line shows the binned mean, and the error bars are the dispersion in each bin.}
\label{fehel}
\end{figure*}

\subsection{The VLT/FLAMES spectroscopic metallicities}

We started by making a reanalysis of the mean metallicity in the Sculptor dSph. The CaT metallicities of individual RGB stars in Sculptor have previously been shown to be reliable as an \feh\ indicator, by comparing them to [\ion{Fe}{I}/H] measurements for the same stars made with high-resolution spectra \citep[e.g.][]{Battaglia08a, Starkenburg10, Hill19}, and we have repeated this exercise in Appendix~C (see Fig. \ref{uves}).

The metallicity distribution function (MDF) of the 1339 \feh\ measurements (S/N$>13$) is shown in Fig.~\ref{fehhist}. A Gaussian is a reasonable match to the distribution. This could be marginally improved for a distribution with two distinct components (metal-poor and metal-rich). This is consistent with two distinct metallicity components \citep[e.g.][]{Tolstoy04, Battaglia08b}. However, it is not clear that any metallicity distribution needs to be Gaussian. In addition, the distinction between two populations and a gradient is not easy to draw. In Fig.~\ref{fehhist} we also show the MDF corrected for the  incompleteness as a function of elliptical radius for the different radial bins as determined in Fig.~\ref{figellcomp} (as an open black histogram and a black dashed line Gaussian fit). 
This assumed that the \gdr{3} RGB members are representative of the mass of stars in the galaxy. In Fig.~\ref{fehhist} it can be seen that the effect is not likely to be large, if this assumption was reasonable. The shift in the mean of the population appears to be small, only $\sim 0.1$~dex.
However, outliers cannot be corrected for as they are rare objects, so the prime reason to increase the completeness of the observations is to look for the true number of rare objects, such as extremely metal-poor stars or carbon-rich stars.

The fact that the central region of Sculptor was found to be significantly more metal-rich on average than the outer regions likely explains the higher average metallicity of Sculptor determined by \cite{Kirby09}, who focused on the inner regions of Sculptor (see Table~\ref{tablepm}). This could also be an offset in the calibration caused by different metallicity indicators used.  It should be noted that in our CaT method metal-rich stars suffered more from low S/N than metal-poor stars, as the strong sky lines next to the middle line (EW2) can be problematic at low S/N and when the line was wide it tended to overlap more with the strong sky line residuals. In all methods lines get weaker as stars get more metal poor.

\begin{figure}
\centering
\includegraphics[width=1.\linewidth]{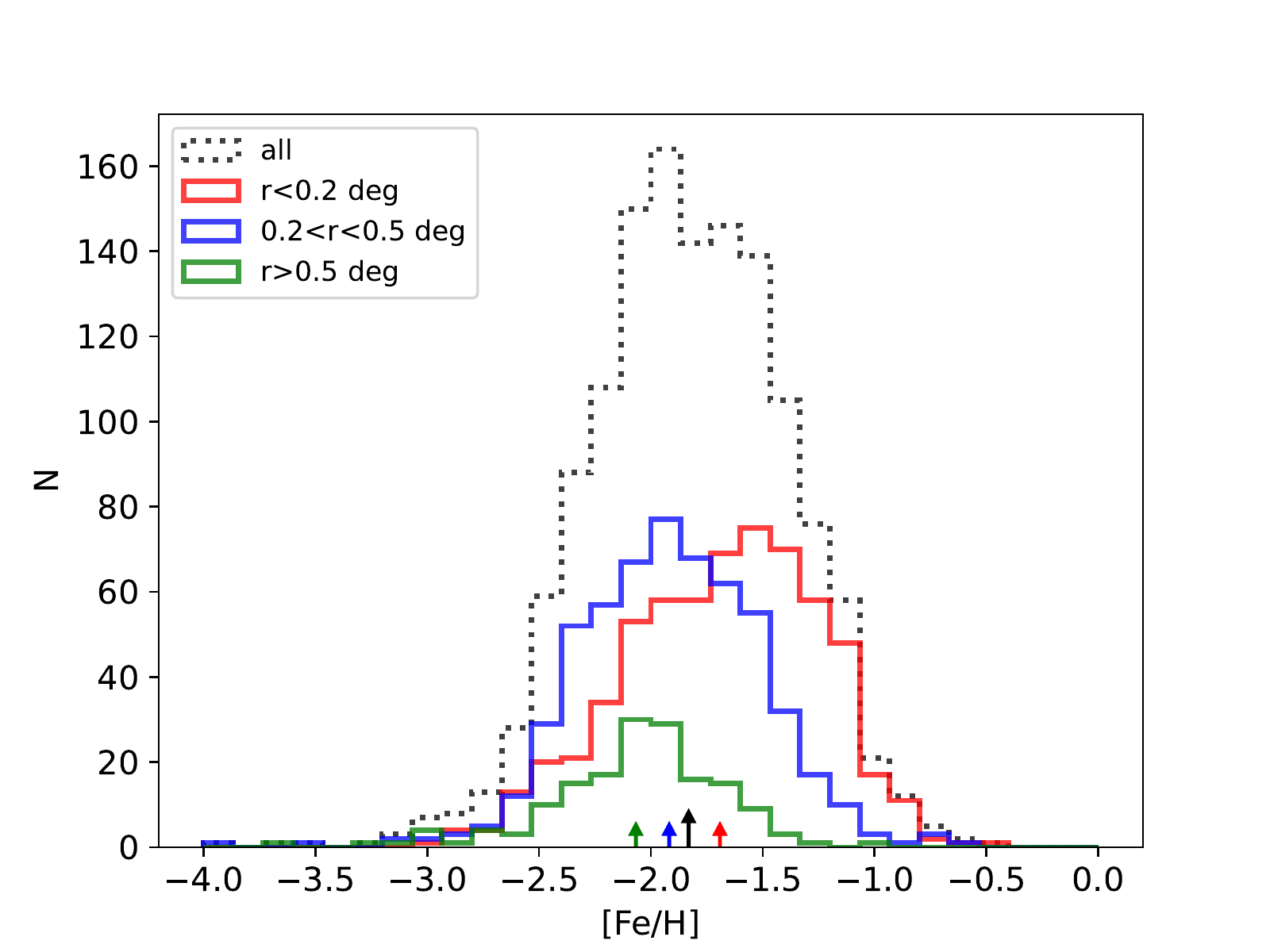}
\caption{VLT/FLAMES metallicities for 1339 RGB member stars at different ranges of elliptical radius (in degrees), r$_{ell}$, in the Sculptor dSph. Coloured arrows show the shift in mean metallicity for the different r$_{ell}$ ranges.} 
\label{fehhistrad}
\end{figure}

\begin{figure}
\centering
\includegraphics[width=0.75\linewidth]{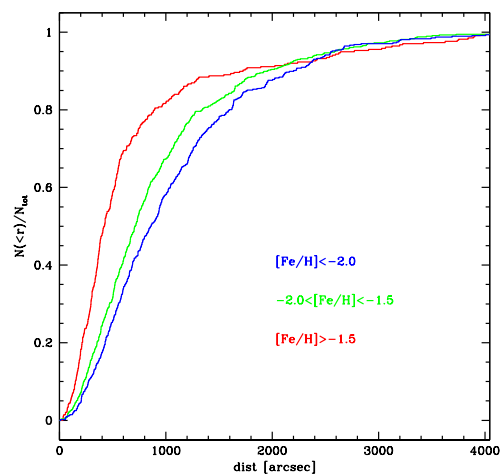}
\caption{Cumulative distributions for three sub-populations of member stars with different metallicities (colours), with increasing distance on the sky from the centre of the Sculptor.}
\label{radpops}
\end{figure} 
 
For the 1339 RGB stars in our sample of VLT/FLAMES LR8 spectra in the Sculptor dSph with accurate \feh, we plot \feh\ as a function of elliptical radii, r$_{ell}$, in Fig.~\ref{fehel}. Here we also plot the binned mean and the dispersion in each bin, which is consistent with a straight line fit.
The clear metallicity gradient is consistent with what has previously been noted \citep[e.g.][]{Tolstoy04, Kirby09}, although the scatter plot is much denser and more extended than previous results, showing that there is a lot of intrinsic scatter in \feh\  at any r$_{ell}$, and the very few points lying at particularly high or low metallicities typically appear to be robust members. For the bulk of the measurements the scatter was much larger than the intrinsic errors on the measurements. This is consistent with a galaxy experiencing extended star formation, as opposed to a limited burst, where less scatter in the metallicities might be expected. The mean metallicity in the central bin of Fig.~\ref{fehel} is \feh=$-$1.65 with a dispersion of $\pm$0.44 and in the outermost region the numbers of stars decrease dramatically, and the outermost bin in Fig.~\ref{fehel} contains four stars, and so the mean \feh=$-$2.47 with a dispersion of $\pm$0.31 is not robust. Enlarging the last bin to outermost 8 or even 24 stars, leads to a mean \feh=$-$2.2 and a dispersion of $\pm$0.36.

There is no clumpiness or any obvious signs of linear features that might suggest streams or groups of stars with similar metallicities and/or \feh\  values that differ from the mean. However, the dynamical timescale of a galaxy like Sculptor is very short ($\sim 50-100$~Myr), and so evidence of past events will quickly disappear. Figure~\ref{fehel} also does not show an obvious presence of two populations, and it seems to be more consistent with a radial gradient and scatter around the typical metallicity at each radius. It still might be that different populations are hidden in the scatter. This is further explored in the next subsection, which includes the kinematics.

\begin{figure}
\centering
\includegraphics[width=1.\linewidth]{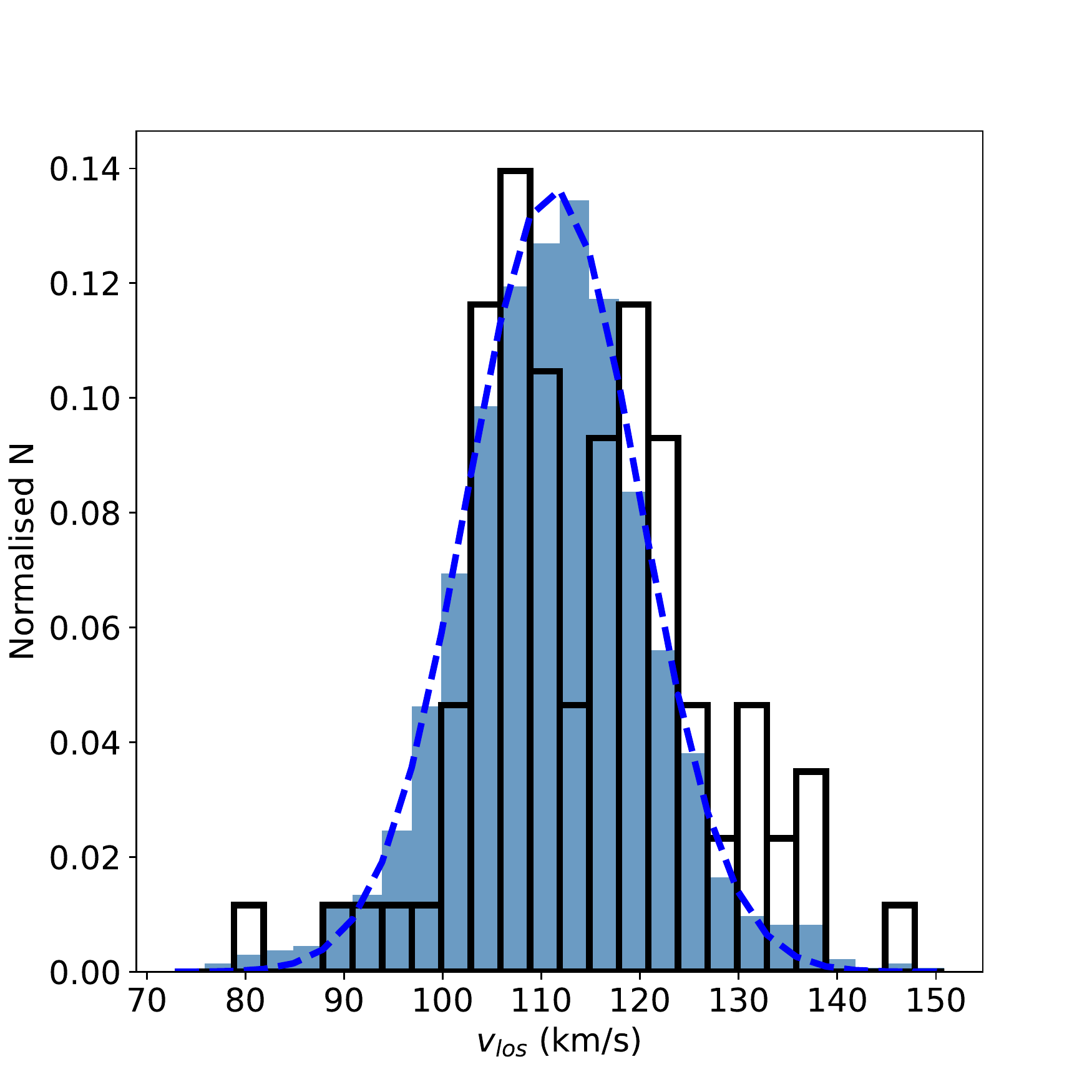}
\caption{ Normalised histogram of VLT/FLAMES \vlos\ measurements (S/N$>$ 13) for the 1339 RGB members of the Sculptor dSph with reliable \feh\ determinations (blue). The shape of the distribution is fit with a Gaussian function (dashed blue line). In addition, the 86 stars with \feh$< -2.5$ are shown as a (black) open  normalised  histogram.}
\label{velhist}
\end{figure}

\begin{figure*}[ht]
\centering
\includegraphics[width=1.\linewidth]{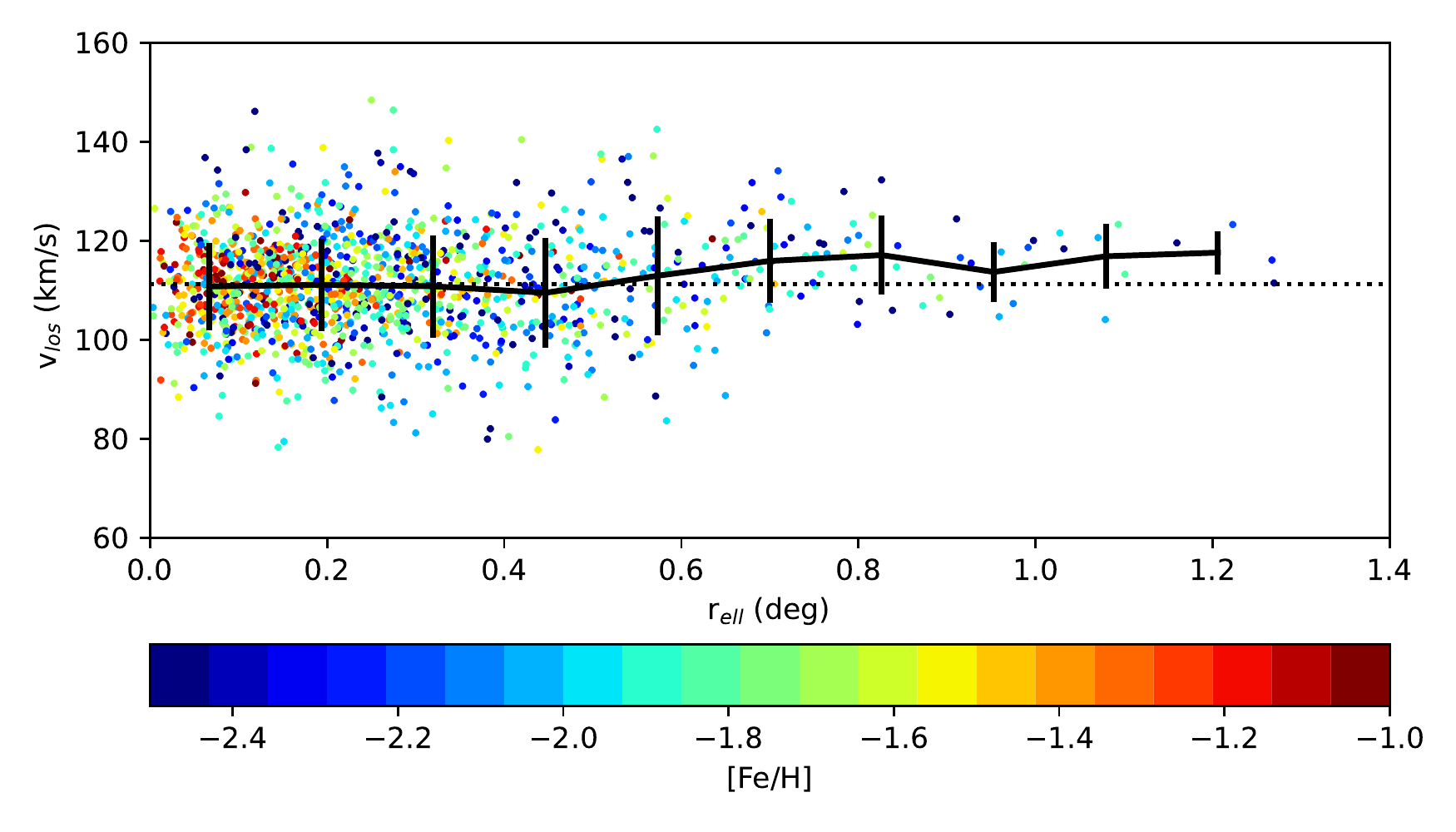}
\caption{ VLT/FLAMES heliocentric los velocities, \vlos, for the 1339 RGB stars in our sample with the most reliable \feh\  measurements as a function of elliptical radius, r$_{ell}$, in the Sculptor dSph.
The dotted black line shows the mean \vlos$=+111.2$~km/s, and the binned mean and the dispersion are plotted as a solid black line and error bars.}
\label{vhels}
\end{figure*}

In general, the metal-poor population is consistently, if sparsely, distributed throughout the galaxy (Fig.~\ref{fehel}), whereas the metal-rich component is clearly more centrally concentrated. This results in a mean metallicity gradient of $-0.7$~dex~deg$^{-1}$. The intrinsic spread in metallicity seems to decrease towards the outer regions of Sculptor, but it is challenging to disentangle this from the even more rapid fall in the number of measurements (representative of the falling stellar density). 
In Fig.~\ref{fehhistrad} we show the metallicity gradient in the form of a histogram for different regions in elliptical radius. In Fig.~\ref{radpops} we show the cumulative radial distributions in three different metallicity ranges, with roughly equal numbers of stars, and it can be seen that the distributions differ, which is representative of the metallicity gradient seen in Figs~\ref{fehel} and~\ref{fehhistrad}. These quantities are often derived by theoretical models and simulations of the stellar populations of dwarf galaxies \citep[e.g.][]{Gelli20}, and can be used as a reference for future comparisons.

\begin{figure*}[ht]
\centering
\includegraphics[width=1.\linewidth]{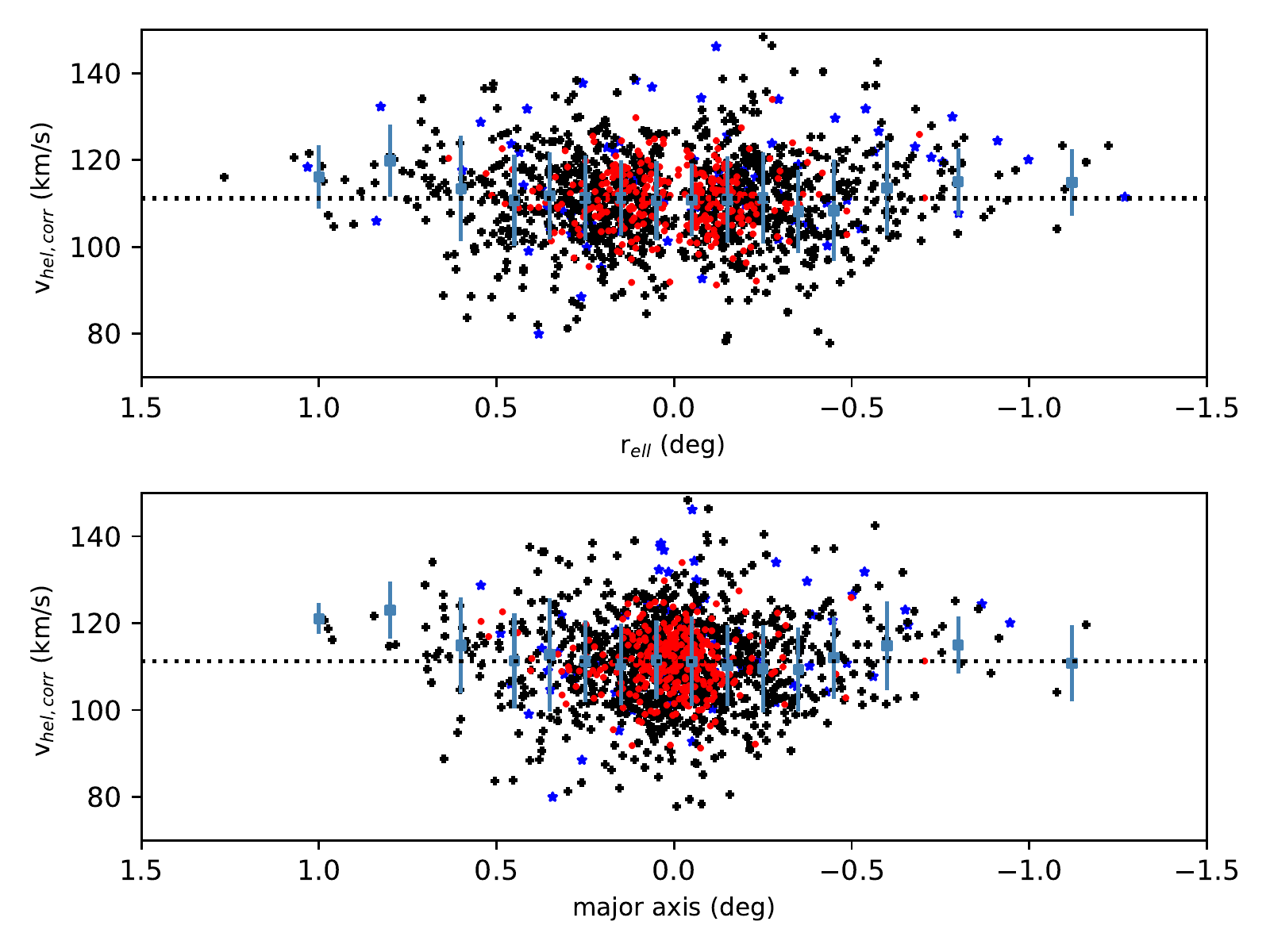}
\caption{ VLT/FLAMES LR \vlos \ corrected for the shift created by the motion of the Sun around the Galactic centre at the position of Sculptor, colour-coded by \feh, as a function of elliptical radius, r$_{ell}$ (top) and the position along the major axis (bottom). 
The colour-code is blue stars: \feh$< -2.5$; black crosses $-2.5\leq$\feh$<-1.5$; and red circles \feh$\geq-1.5$. 
Binned mean velocities are shown as steel blue points, with their standard deviation as bars. } 
\label{vgsr}
\end{figure*}

\subsection{The combined kinematic and metallicity properties}

By adding \vlos\ to the \feh\ measurements we can look at the chemo-dynamical properties of the Sculptor dSph. The normalised histogram of the 1604 \vlos\ measurements (S/N$>7$) in our new sample is shown in Fig.~\ref{velhist}, along with a Gaussian fit and a vertical line at the mean systemic velocity (111.2\,km/s; see Table~\ref{tablepm}). This looks to be a normal distribution with the possible exception of a higher number of stars at larger \vlos. The tail is not a feature of low S/N measurements, as it remains even if only the high S/N ($>13$) measurements are included.

Also plotted in Fig.~\ref{velhist} is the normalised histogram distribution of  the 86 stars with \feh~$\leq-2.5$. It clearly looks different the total distribution, falling into two peaks and extending very clearly to higher velocities.
A Kolmogorov-Smirnov test was carried out, for the metallicity selections given in Table~\ref{tablepm} and they all tend to be similar to each other, except at \feh$<-2.2$, where the difference starts to increase, and is significant only for \feh$<-2.5$. The stars with \feh$<-2.5$ have the most clear offset in mean velocity, partly because they are no longer a Gaussian distribution. This is barely significant given the errors and the small number of stars in this metallicity range, but it is a hint that the kinematics of the most metal population may be different and this should be followed up with larger samples.

\begin{figure*}[ht]
\centering
\includegraphics[width=1.\linewidth]{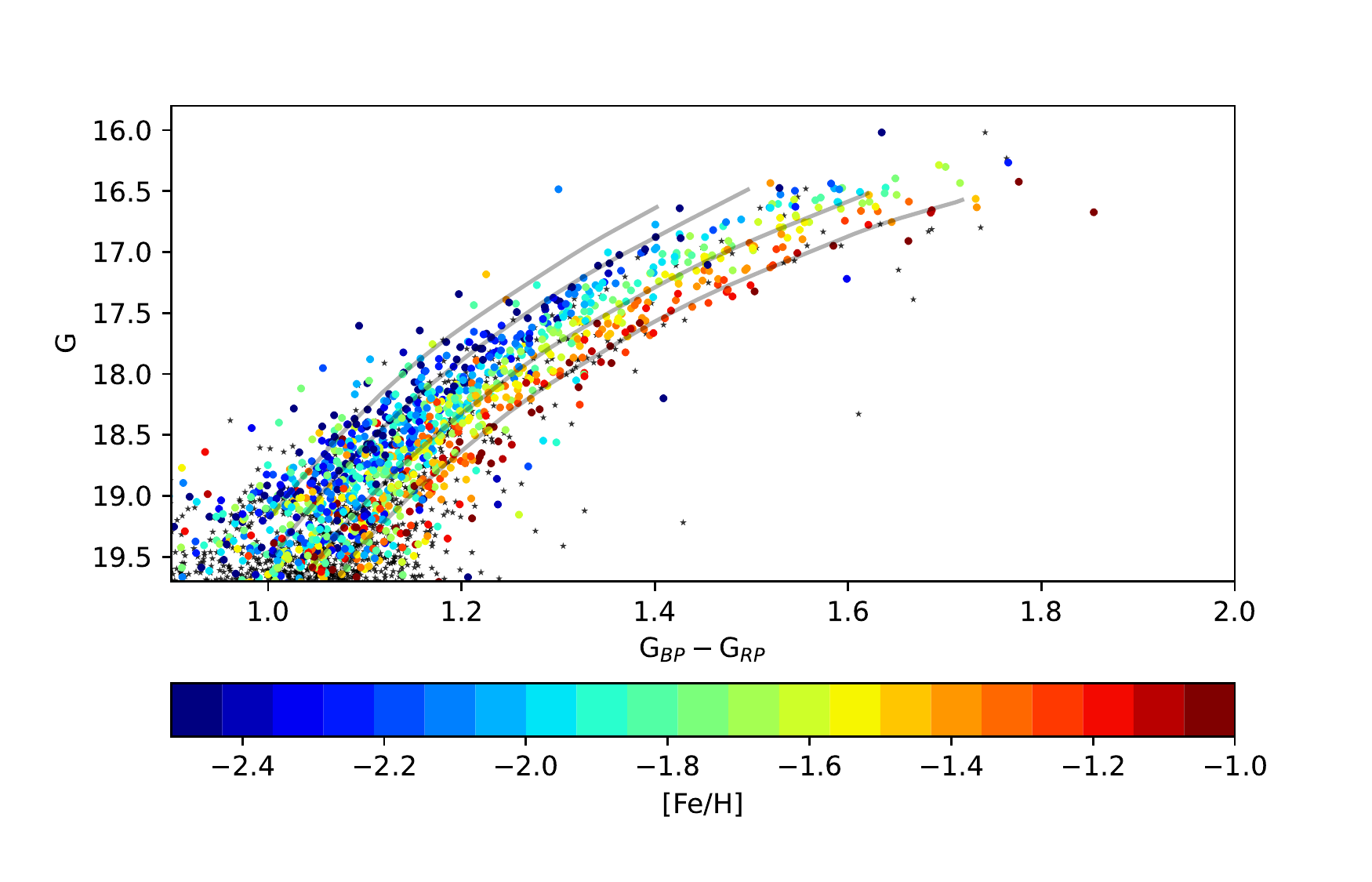}
\caption{
 \gdr{3} CMD for the{ 1339} (S/N$>13$) members of the Sculptor dSph, colour-coded by \feh. Small black stars are \gdr{3}\ members without VLT/FLAMES spectroscopy or insufficiently accurate \feh\  measurements. Grey lines are BASTI RGB isochrones, with [$\alpha$/Fe]$=+0.4$, for (from left to right):  $\rm\feh=-3.2,-2.2, -1.7, -1.4$; and $\rm age=13, 13, 12, 10$\,Gyr.}
\label{cmd2}
\end{figure*}

\begin{figure*}[ht]
\centering
\includegraphics[width=1.\linewidth]{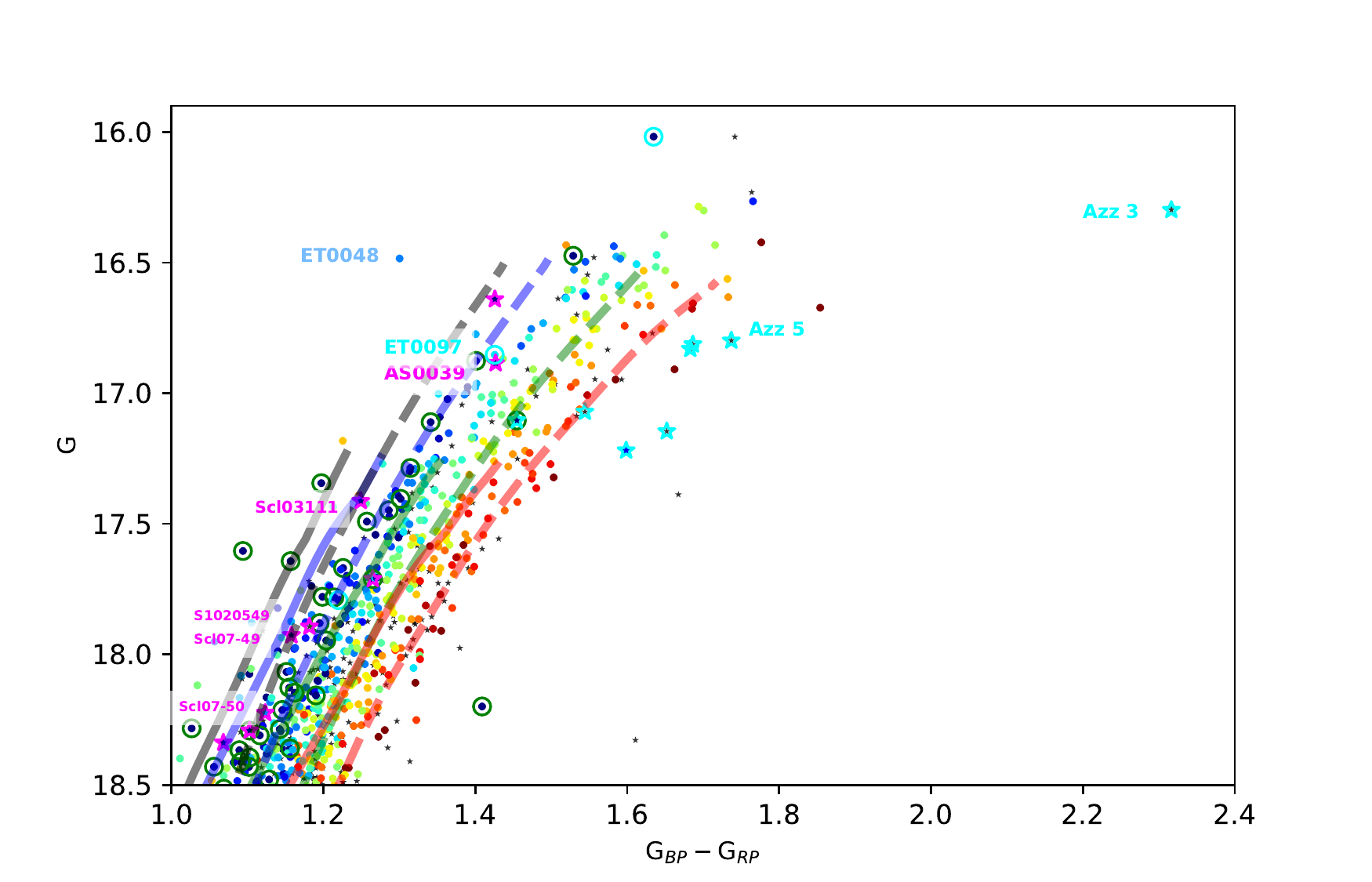}
\caption{
Upper \gdr{3} CMD for all members of the Sculptor dSph, colour-coded by \feh\  as in Fig.~\ref{cmd2}. The eight cyan stars symbols come from the carbon star catalogue of \citet{Azz85}, and the three open cyan circles are C-rich stars from later studies. Magenta star symbols are stars with \feh$<-3$ in our sample, while green circles are those with $-3<\feh<-2.5$. BASTI isochrones from Fig.~\ref{cmd2} are in dashed coloured lines, with the early AGB evolutionary phase shown as solid lines.
}
\label{cmdnam}
\end{figure*}

To investigate the velocity and metallicity distribution in more detail, we plot the velocities of individual member stars against elliptical radius (r$_{ell}$), in Fig.~\ref{vhels}, colour-coded by CaT metallicity (\feh). It can be clearly seen, as has already been established, that the metal-rich population is more centrally concentrated than the 
metal-poor population, and also has a lower velocity dispersion  \citep{Tolstoy04, Battaglia08b}. The Sculptor stars withing the central region are now better studied, and the metal-rich population (\feh$> -1.7$) is shown to extend further out in the galaxy than previous smaller samples suggested.
The highest metallicity stars show a marked central concentration, limited to r$_{ell}\lesssim0.5$\,deg. In addition, the velocity dispersion declines slightly moving from the centre of Sculptor ($\sim$9km/s) to the outermost regions with a dispersion of $\sim$4$-$6km/s. In addition the central velocity also changes from 110.9~km/s in the inner regions to $116.5-117.5$~km/s in the outer regions. 

This new much larger and more accurate dataset appears to lack the rotation signature seen by \citep{Battaglia08b}, compatible with the results by \citet{MG23}.
There is a slight gradient in the mean los velocity, $\langle\vlos\rangle$, moving from the centre of Sculptor outwards. This is not an effect of the 
bulk motion of the galaxy projected onto the proper motion and/or radial velocity at different positions in the galaxy \citep{Feast61,Feitzinger77,Meatheringham88,KapStrig08}, as
assuming the proper motion given by \citep{Battaglia22}, and the distance from \citep{MV15}, the tangential motion of Sculptor is 74.4~km/s, and so 
the most significant \vlos difference is 1.3 km/s [$ \sin( \rho)$], and this only leads to an apparent velocity gradient of 1.2 km/s/deg, which is not enough to explain what is seen in Fig.~\ref{vhels}. The 
rotation of the Sun around the Milky Way compared to the position of Sculptor dSph on the sky results in a larger effect. In Fig.~\ref{vgsr} we show the \vlos corrected for the Galactic standard of rest, using the Gala python routine provided by Adrian M. Price-Whelan\footnote{ \tiny \url{http://gala.adrian.pw/en/latest/api/gala.coordinates.vhel_to_vgsr.html }}. 
The mean velocities are plotted and the standard deviation is shown as a range of velocity. The mean \vlos\ of stars in Sculptor increases by 6\,km/s from the central region to the outskirts, giving a gradient of around +5\,km/s\,deg$^{-1}$. Figure~\ref{vgsr} suggests that this is a fairly symmetric effect from one side of the galaxy to the other, along the major axis. Changing the area around the major axis that is included in the plot makes no difference to the gradient. It can also been seen from the direction of the proper motion on the sky, given as a blue arrow in Fig.\ref{fig:flamesmemskycmd}a, that the elongation of Sculptor is not along the direction of motion.
However, the number of stars in the outer regions is relatively small, so the effect is not highly significant but it can be seen in both Fig.~\ref{vhels} and Fig.~\ref{vgsr} that beyond $r_{ell} > 0.6$~deg the stars tend to have velocities higher than the systemic velocity. This could perhaps suggest that the outer regions of the galaxy are more loosely bound than the centre as might be expected for a small galaxy moving around in the tidal field of a much larger galaxy. There are no obvious clumpy velocity or velocity-metallicity structures, which could have suggested that the presence of dissolving globular clusters or other mergers events. It has been shown that interpreting the velocity structure in the outer regions of small dwarf galaxies orbiting in the potential of the Milky Way is not straightforward \citep{vdBosch18}. It is clear that to fully understand the complex dynamics of Sculptor  careful dynamical modelling of this dataset will be needed. This is  beyond the scope of this paper, but re-assessing in detail the internal chemo-dynamical properties of Sculptor will be the subject of future work.

\subsection{The colour-magnitude diagram}

In addition to the exquisite astrometry, the \gaia\ catalogue also includes uniquely accurate and well calibrated photometry in 3 photometric bands: G, G$_{BP}$ and G$_{RP}$ \citep{Evans18, Riello21}. The depth of the \gaia\ photometry for the Sculptor dSph (Fig.~\ref{fig:flamesmemskycmd}c, d) is quite a bit less than what has been possible from deep ground-based surveys \citep[e.g.][]{deBoer11, deBoer12,Bettinelli19}, and certainly the \textit{Hubble} Space Telescope \citep{Dolphin02}. However, the \gaia\ photometry is extremely well matched to our spectroscopic survey of the RGB, and it has the advantage of uniformly covering the whole galaxy on the sky. 

One of the most stunning results comes from the excellence of the \gaia\ photometry. A clear metallicity gradient is seen across the RGB  (Fig.~\ref{cmd2}), where the metal-rich and metal-poor populations lie along distinct isochrones, with the metallicity increasing with { $(\gbp-\grp)$} colour.
This is a much clearer distinction than has been seen before and is only possible because of the accurate and extensive \gdr{3} photometry, combined with the uniquely accurate membership determinations from \gdr{3} astrometry and VLT/FLAMES los velocities. Plotted on top of the CMD in Fig.~\ref{cmd2} are four isochrones taken from the $\alpha$-rich ([$\alpha/\mathrm{Fe}] = +0.4$) a Bag of Stellar Tracks and Isochrones (BASTI) set \citep{Hidalgo18, Pietrinferni21}. The choice of isochrones is informed by the SFH of Sculptor \citep{deBoer12, Savino18, Bettinelli19}, and high-resolution spectroscopic abundances \citep{Hill19}. It is interesting to see the age-metallicity relation so clearly in the data and also how well { they match} the shape as well as the position of the isochrones. This comparison assumes a distance modulus, $m-M=19.62$ and a reddening of $E(B-V)=0.018$ \citep{Battaglia22}. Arguably the most metal-rich isochrone ($[\mathrm{Fe}/\mathrm{H}]=-1.4$) should be solar-scaled ([$\alpha/\mathrm{Fe}]=0$), as high-resolution spectroscopy in the central region shows that at this \feh\ the stars typically have [$\alpha/\mathrm{Fe}]\approx0$. However, this isochrone ($[\mathrm{Fe}/\mathrm{H}]=-1.4$, [$\alpha/\mathrm{Fe}]=0$) did not fit the stellar distribution at all, even when increasing the age to 13\,Gyr. A detailed analysis of the CMD is left to future work.

On the red side of the RGB are typically Carbon-rich stars (Fig.~\ref{cmdnam}), many of which have been first identified by \citet{Azz85}. They are likely to be dominated by stars that have been enriched by interactions with a binary companion. Intrinsically carbon-enhanced metal-poor (CEMP) stars are more likely to be found more on the blue (metal-poor) side of the RGB. For example, the CEMP-no star found by \citet{Skuladottir15a}, ET0097, is on the blue edge of the RGB (see Fig.~\ref{cmdnam}). On the blue side of the RGB is where the most metal-poor stars are to be found, as is shown in Fig.~\ref{cmd2}. The scatter far from the RGB is most likely due to stars following unusual evolutionary tracks (e.g. ET0048 from \citealt{Hill19}); they could be "rejuvenated" blue-straggler type objects, of which there are quite a few on the main sequence (e.g. \citealt{Mapelli09}), where they stand out more clearly. There are certainly ancient asymptotic giant branch (AGB) stars in the Sculptor \gdr{3} CMD. Looking at dedicated AGB isochrones \citep[e.g.][]{Marigo13} to fit the outliers beside and above the RGB requires extremely young ages ($\sim 2$--$4$~Gyr), which are clearly inconsistent with the SFH determined from the much more reliable and much more populated main sequence turnoff region \citep[e.g.][]{deBoer12}. Thus if these stars are AGB stars, and thus members of Sculptor, they are likely to be stars that have been rejuvenated by binary interactions. In Fig.~\ref{cmdnam}, for consistency we use the BASTI isochrones, which show the early AGB phase. It is clear that they overlap with the RGB, especially on the blue side: younger, more metal-rich AGB stars are likely to overlap with older, more metal-poor RGB stars. Dedicated AGB models \citep{Marigo13} go to higher luminosities, up to and above the tip of the RGB and might explain more of the scatter here in Fig.~\ref{cmdnam}. 
Picking out the AGB stars is important for the purposes of understanding the spectroscopic results in terms of chemical evolution and SFH. The AGB phase can potentially explain some of the metal-poor stars bluer than the bulk of the RGB. The distinction between an AGB and an RGB star is difficult to make observationally, even on the basis of HR spectroscopy. However, they are expected to have excess carbon in their spectra due to the various dredge up and mixing processes that occur during the AGB stellar evolution phase. A detailed analysis of these evolved stars in Sculptor is left to a future work.

\subsection{The most metal-poor stars}

For the past two decades, astronomers have searched the Sculptor dSph looking for extremely metal-poor stars \citep[e.g.][]{Tolstoy03, Tafelmeyer10,Frebel10,Starkenburg13,Jablonka15, Simon15, Skuladottir15a, Skuladottir21, Chiti18, Hill19}. These searches have been less fruitful than originally expected, and although there are a couple of stars around $\feh\sim -4$, the extremely metal-poor tail ($\feh<-3$) is not well populated. Another interesting aspect is the lack of CEMP stars without s-process enhancement (CEMP-no). These are progressively more common in the Galactic halo, and in ultra-faint dwarf galaxies  at $\feh<-2$. There are one or two unambiguous examples of these stars in the Sculptor dSph, a conspicuous absence compared to the Milky Way and ultra-faint dwarfs (e.g.~\citealt{Skuladottir15a, Salvadori15, Kirby15}). Of the known sample of C-rich stars from \citet{Azz85}, two appear to be metal poor, and could possibly be CEMP-no stars. The rest are more likely to be highly evolved carbon stars or CEMP-s stars, but limited high-resolution spectroscopy has been carried out on this sample due to the complexities of dealing with strong C features in the spectra. In some of the more extreme cases it is even challenging to accurately determine \feh.

In Fig.~\ref{cmdnam} we show in the CMD, both extremely metal-poor ($\feh<-3$) RGB stars in our Sculptor VLT/FLAMES LR8 sample, and also where C-rich stars have been found. Most of the bright VLT/FLAMES targets with $\feh<-2.5$ have already been observed with follow-up intermediate- to high-resolution spectra, although in a few cases the analysis is still pending. 
In Fig.~\ref{cmdnam} we also identify some of the stars that have been followed up with high resolution spectroscopic abundance analysis. Of the stars with metallicity $\feh<-3$, Scl03111 ($\feh=-3.61$) is from \citet{Jablonka15}, and S1020549 is from \citet{Frebel10}, originally given as $\feh=-3.81$, but reanalysed by \citet{Simon15} to have $\feh=-3.68$. Scl07-50 is the most metal-poor star in the \citet{Tafelmeyer10} sample, at $\feh=-3.96$, but reanalysed by Simon et al. to have $\feh=-4.05$. There is also Scl07-49, analysed by Tafelmeyer et al. to have $\feh=-3.48$, and re-analysed by Simon et al. to be $\feh=-3.28$. The star AS0039 has been analysed by \citet{Skuladottir21} to have $\feh=-4.11$, and is thus currently the most metal-poor star known in Sculptor dSph.  The most metal-poor isochrones match the positions of these stars quite well, and none of them appear to be an obvious AGB candidate, given how metal-poor they are. Only some of the stars further to the blue from the main RGB or above the tip of the RGB are  likely to be AGB stars. A more detailed analysis of the stars that do not match the isochrone positions is needed to see if there are AGB stars buried in the RGB population. This distinction is not easy to make on the basis of spectroscopy.

\section{Conclusions}

With a combination of new and old re-processed archival VLT/FLAMES LR8 data, we have obtained a uniform sample of \feh, \vlos, and \gdr{3} astrometric parameters for 1604 individual member RGB stars. This is the largest sample of individual measurements made to date for a classical dSph galaxy, and they extend from the inner regions to the outermost. This has allowed us to look again at the chemo-dynamical properties of the Sculptor dSph. The main difference with previous results is that, because of the much larger, more uniformly distributed, and more precise `cleaner' sample, we see more members of different sub-populations; thus the scatter is much better quantified and the differences become subtler. This will allow more sophisticated and detailed modelling of this galaxy.

We confirm that the Sculptor dSph has complex \vlos \ and \feh\  distributions. We present a new and more complete MDF for Sculptor, including a correction for the incompleteness of the spectroscopic data. We no longer find the rotation signature from earlier work with part of this sample. The kinematic differences between metal-rich and metal-poor populations appear more intricate than originally thought, with the strongest differences in the global properties being found in this analysis for the 86 most metal-poor stars. The wealth of information coming from the combination of spectroscopic metallicities and the exquisite photometry of the Sculptor dSph RGB reveals the complex buildup of metals and the subtler and more difficult-to-interpret motions of the stars within the galaxy.

\section*{Acknowledgements}

This work has made use of data from the European Space Agency (ESA) mission
\gaia\ (\url{https://www.cosmos.esa.int/gaia}), processed by the \gaia\
Data Processing and Analysis Consortium (DPAC,
\url{https://www.cosmos.esa.int/web/gaia/dpac/consortium}). Funding for the DPAC
has been provided by national institutions, in particular the institutions
participating in the \gaia\ Multilateral Agreement.
This project has received funding from the European Research Council (ERC) under the European Union’s Horizon 2020 research and innovation programme (grant agreement No. 804240) for S.S. and Á.S. G.B. acknowledges support from the Agencia Estatal de Investigación del Ministerio de Ciencia en Innovación (AEI-MICIN) and the European Regional Development Fund (ERDF) under grant number AYA2017-89076-P, the AEI under grant number CEX2019-000920-S and the AEI-MICIN under grant number PID2020-118778GB-I00/10.13039/501100011033. P.J. acknowledges support from the Swiss National Science Foundation. E.T. thanks Filippo Fraternali for interesting discussions about the dynamical properties of dwarf galaxies. We thank the anonymous referee for a careful reading of the manuscript.

\bibliographystyle{aa}
\bibliography{et}

\clearpage
\onecolumn

\begin{appendix}
\renewcommand{\thesection}{A}
\section{The Observations: Individual VLT/FLAMES pointings} 
\label{app:results}
\setcounter{table}{0}
\renewcommand{\thetable}{A.\arabic{table}}

\setlength\LTleft{0pt}            
\setlength\LTright{0pt}           
\begin{longtable}{@{\extracolsep{1pt}}|l|c|l|c|c|c|c|c|c|c|c|}
\caption{\label{tab:obs}Sculptor LR8 observations with VLT/FLAMES/GIRAFFE.}\\
\endfirsthead

\multicolumn{11}{l}%
{\tablename\ \thetable\ -- \textit{Continued from previous page}} \\

No. & Date & Name & Prog ID  & Centre of field & airm & DIMM &exp & spec & vmem & vshift \\\hline
\endhead
\hline \multicolumn{11}{r}{\textit{Continued on next page}} \\
\endfoot
\hline
\endlastfoot
\hline
No. & Date & Name & Prog ID  & Centre of field & airm & DIMM &exp & spec & vmem & vshift \\\hline
1& 2003.09.29 & Scl02a   & 171.B-0588 &  00 58 22.4 ~~ -33 42 36.0 & 1.52 & 0.95  & 1800 & 104 & 63 & 1.125\\
-& 2003.09.29 & Scl02b & 171.B-0588 &  00 58 22.4 ~~ -33 42 36.0 & 1.32 & 0.99  & 1800 & 99  & 60 & 0.821\\
2& 2003.09.29 & Scl11a   & 171.B-0588 &  00 56 08.9 ~~ -33 42 29.0 & 2.12 & 1.06  & 1200 & 41  & 14 & 0.544 \\
-& 2003.09.29 & Scl11b   & 171.B-0588 &  00 56 08.9 ~~ -33 42 29.0 & 1.80 & 1.02  & 1800 & 55  & 22 & 0.658 \\
-& 2003.09.30 & Scl11c   & 171.B-0588 &  00 56 09.0 ~~ -33 42 29.0 & 1.13& 0.60   & 1800 & 54  & 18 & 1.404 \\
3& 2003.09.30 & Scl04a   & 171.B-0588 &  00 59 55.4 ~~ -33 15 53.8 & 1.92 & 0.79  & 1800 & 68  & 16 &0.764 \\
-& 2003.09.30 & Scl04b   & 171.B-0588 &  00 59 55.4 ~~ -33 15 53.8 & 1.58 & 0.61  & 1800 & 73  & 20 & 0.423\\
4& 2003.09.30 & Scl07   & 171.B-0588 &  01 00 10.9 ~~ -34 06 04.0 & 1.37 & 0.74  & 3600 & 76  & 38 & 0.541\\
5& 2003.10.01 & Scl03a   & 171.B-0588 &  01 01 44.1 ~~ -33 43 01.5 & 1.32 & 0.60  & 1800 & 113 & 83 & 0.92\\
-& 2003.10.01 & Scl03b   & 171.B-0588 &  01 01 44.1 ~~ -33 43 01.5 & 1.21 & 0.55  & 1800 & 113 & 78 & 0.916 \\
6& 2003.10.01 & Scl05a   & 171.B-0588 &  00 58 08.8 ~~ -33 10 42.3 & 1.83 & --    & 1800 & 11  &  0 & 0.771\\
-& 2003.10.01 & Scl05b   & 171.B-0588 &  00 58 08.8 ~~ -33 10 42.3 & 1.49 & 0.58  & 1800 & 45  & 14 & 0.555 \\
7& 2003.10.02 & Scl025a  & 171.B-0588 &  00 59 46.4 ~~ -32 40 00.7 & 2.20 & 0.51  & 1200 & 22  &  4 & 1.012\\
-& 2003.10.02 & Scl025b & 171.B-0588 &  00 59 46.4 ~~ -32 40 00.7 & 1.84 & 0.69  & 1554 & 25  &  4 & 1.061\\\hline
8& 2004.08.14 & Scl14a   & 171.B-0588 &  01 03 29.6 ~~ -33 40 47.2 & 1.09 & 0.44  & 3600 & 52  & 19 &0.793\\
-& 2004.10.12 & Scl14b   & 171.B-0588 &  01 03 29.6 ~~ -33 40 47.2 & 1.05 & 0.60  & 3600 & 52  & 24 & 0.791\\
9& 2004.10.11 & Scl20   & 171.B-0588 &  01 06 14.5 ~~ -33 41 23.4 & 1.08 & 0.90  & 3600 & 33  &  4& 1.146 \\\hline
10& 2004.09.10 & SclCen  & 072.D-0245 &  01 00 03.8 ~~ -33 41 29.7 & 1.08 & 0.86 & 3600 & 114 & 99 & 0.439 \\\hline
11& 2005.11.07 & Scl103a & 076.B-0391 &  00 57 24.8 ~~ -33 34 37.7 & 1.18 & 0.68 & 3600 & 71  & 45 & 0.941\\
-& 2005.11.07 & Scl103b & 076.B-0391 &  00 57 24.8 ~~ -33 34 37.7 & 1.01 & 0.92  & 3600 & 71  & 51 & 0.928 \\
12& 2005.11.07 & Scl109 & 076.B-0391 & 01 01 11.3 ~~ -33 56 50.8 & 1.04 & 0.60   & 3600 & 98  & 80 & 1.001\\
13& 2005.11.07 & Scl106 & 076.B-0391 & 01 01 23.7 ~~ -33 23 39.7 & 1.15 & 0.68   & 3600 & 85  & 50 & 0.269\\
14& 2005.11.07 & Scl101 & 076.B-0391 & 01 02 52.3 ~~ -33 52 06.0 & 1.06 & 1.13   & 3600 & 52  & 19 & 1.153 \\
15& 2005.11.08 & Scl108a  & 076.B-0391 & 00 58 59.9 ~~ -33 27 04.9 & 1.13 & 1.21  & 2700 & 95  & 70 & 0.736\\
-& 2005.11.08 & Scl108b  & 076.B-0391 & 00 58 59.9 ~~ -33 27 04.9 & 1.25 & 1.01   & 2700 & 95  & 48 & 0.916 \\\hline
16& 2007.06.21 & Scl207  & 079.B-0435 & 01 03 52.5 ~~ -34 12 59.8 & 1.22 & 1.05  & 3600 & 37  & 14 & 3.065\\
17& 2007.06.21 & Scl203  & 079.B-0435 & 01 04 35.0 ~~ -33 24 40.2 & 1.53 & 0.78  & 3600 & 31  &  7 & 0.207\\
18& 2007.06.21 & Scl200a  & 079.B-0435 & 01 05 10.4 ~~ -33 49 04.9 & 1.12 & 0.88  & 3600 & 36  & 10 & -3.338\\
-& 2007.06.21 & Scl200b  & 079.B-0435 & 01 05 10.4 ~~ -33 49 04.9 & 1.03 & 1.08   & 1800 & 36  & 8 & -3.448 \\
19& 2007.08.14 & Scl202a  & 079.B-0435 & 00 54 23.1 ~~ -33 31 44.9 & 1.01 & 0.54  & 1800 & 28  &  3 & -0.128\\
- & 2007.09.14 & Scl202b  & 079.B-0435 & 00 54 23.1 ~~ -33 31 44.9 & 1.22 & 0.94  & 3600 & 30  &  7 & 0.227\\
20& 2007.09.12 & Scl220  & 079.B-0435 & 01 03 13.2 ~~ -33 13 01.5 & 1.11 & 1.22  & 3600 & 46  &  12 & 0.834\\
21& 2007.09.13 & Scl204a  & 079.B-0435 & 01 00 38.2 ~~ -32 59 40.1 & 1.11 & --    & 3600 & 30  & 3 & -3.324\\
-& 2007.09.14 & Scl204b  & 079.B-0435 & 01 00 38.2 ~~ -32 59 40.1 & 1.02 & 0.67   & 1800 & 32  &  4  & 0.416 \\
22& 2007.09.14 & Scl211  & 079.B-0435 & 00 56 12.6 ~~ -34 02 10.6 & 1.07 & 1.38  & 3600 & 42  & 6 & -8.377\\
23& 2007.09.14 & Scl209  & 079.B-0435 & 00 57 48.8 ~~ -34 18 41.2 & 1.13 & 1.23  & 3600 & 41  & 12 & -0.280\\
24& 2007.09.14 & Scl206a  & 079.B-0435 & 00 59 34.4 ~~ -34 21 34.6 & 1.21 & 1.61  & 1800 & 48  & 7 & -0.388\\
-& 2007.09.15 & Scl206b  & 079.B-0435 & 00 59 34.4 ~~ -34 21 34.6 & 1.03 & 1.11   & 3600 & 48  & 11 & -7.877\\
25& 2007.09.15 & Scl205  & 079.B-0435 & 01 01 57.6 ~~ -34 17 39.2 & 1.02 & 1.25  & 3600 & 54  &  9 & -0.183\\
26& 2007.09.16 & Scl208  & 079.B-0435 & 00 58 50.9 ~~ -33 58 15.9 & 1.08 & 1.05  & 3600 & 97  & 63 & -8.264\\
27& 2007.12.04 & Scl201  & 079.B-0435 & 00 55 59.5 ~~ -33 13 31.1 & 1.28 & 1.88  & 3600 & 37  &  5 & -1.870\\\hline
28& 2014.07.20 & SclB1 & 593.D-0309 & 00 59 27.0 ~~ -33 37 47.4 & 1.24 & 0.76    & 2700 & 96  & 96 & 0.071\\
-& 2014.07.21 & SclB2 & 593.D-0309 & 00 59 27.0 ~~ -33 37 47.4 & 1.14 & 0.81     & 2700 & 96  & 96 & -0.204 \\
-& 2014.07.30 & SclB3 & 593.D-0309 & 00 59 27.0 ~~ -33 37 47.4 & 1.38 & 0.68     & 2700 & 96  & 96 & 2.903 \\
-& 2014.08.16 & SclB4 & 593.D-0309 & 00 59 27.0 ~~ -33 37 47.4 & 1.25 & 0.71     & 2700 & 96  & 96 & 0.232 \\
-& 2014.08.19 & SclB5 & 593.D-0309 & 00 59 27.0 ~~ -33 37 47.4 & 1.16 & 1.35     & 2700 & 96  & 96 & 0.811 \\
-& 2014.10.24 & SclB6 & 593.D-0309 & 00 59 27.0 ~~ -33 37 47.4 & 1.15 & 1.03     & 2700 & 96  & 96 & 0.323 \\
-& 2015.07.18 & SclB7 & 593.D-0309 & 00 59 27.0 ~~ -33 37 47.4 & 1.06 & 1.31     & 2700 & 96  & 96 & 1.575 \\
-& 2015.10.04 & SclB8 & 593.D-0309 & 00 59 27.0 ~~ -33 37 47.4 & 1.01 & 0.94     & 2700 & 96  & 96 & 2.166 \\\hline
29& 2016.10.21 & SclB9 & 098.D-0160 & 00 59 27.0 ~~ -33 37 47.4 & 1.50 & 0.96     & 2700 & 98  & 97 & 1.274 \\\hline
30 & 2017.10.13 & SclPM & 0100.B-0337 & 01 00 05.5 ~~ -33 50 32.5 & 1.02 & 0.56  & 3600 & 115 &112 & 2.134\\
31 & 2017.10.13 & SclCen3 & 0100.B-0337 & 01 00 13.4 ~~ -33 40 45.3 & 1.23 & 0.45& 3600 & 114 &112 & 1.471\\
-  & 2017.10.13 & SclCen2 & 0100.B-0337 & 01 00 13.4 ~~ -33 40 45.3 & 1.08 & 0.46& 3600 & 115 &113 & 1.646\\
32 & 2017.10.25 & SclOut1 & 0100.B-0337 & 01 01 19.5 ~~ -33 42 58.5 & 1.04 & 0.78& 3600 & 66  & 62 & 2.100 \\\hline
33 & 2018.09.09 & SclC01 & 0101.D-0210 & 01 00 10.5 ~~ -33 50 42.0 & 1.03 & 1.30 & 3600 & 106 &105 & 1.458\\
34 & 2018.09.09 & SclC02 & 0101.D-0210 & 01 00 10.6 ~~ -33 36 41.5 & 1.14 & 1.00 & 4200 & 105 &104 & 1.053 \\
35 & 2018.09.17 & SclC12a & 0101.B-0189 & 01 00 00.0 ~~ -33 32 30.4 & 1.02 & 0.52 & 1700 & 85  & 52 & 1.272\\
- & 2018.09.17 & SclC12b & 0101.B-0189 & 01 00 00.0 ~~ -33 32 30.4 & 1.01 & 0.66  & 3600 & 92  & 91 & 1.200 \\
36 & 2018.09.17 & SclC14 & 0101.B-0189 & 00 58 59.2 ~~ -33 49 20.0 & 1.05 & 0.72 & 4100 & 92  & 91 & 1.292 \\
37 & 2018.09.17 & SclC13 & 0101.B-0189 & 01 00 09.4 ~~ -33 50 19.6 & 1.20 & 0.64 & 4100 & 72  & 69 & 0.959\\
38 & 2018.09.17 & SclC11 & 0101.B-0189 & 01 02 31.0 ~~ -33 38 49.4 & 1.16 & 0.52 & 3600 & 60  & 58 & 1.509 \\
39 & 2018.09.18 & SclC17 & 0101.B-0189 & 01 00 07.0 ~~ -33 44 11.4 & 1.06 & 1.26 & 4100 & 103 & 93 & 1.626 \\
40 & 2018.09.19 & SclC15 & 0101.B-0189 & 01 00 18.1 ~~ -33 45 59.1 & 1.06 & 0.63& 4100 & 88  & 71 & 1.439\\
41 & 2018.09.19 & SclC18 & 0101.B-0189 & 01 00 48.1 ~~ -34 10 58.9 & 1.22 & 0.78 & 3600 & 26  & 23 & 1.510 \\
42 & 2018.09.19 & SclC10 & 0101.B-0189 & 00 58 46.1 ~~ -33 05 29.4 & 1.04 & 0.56 & 3600 & 15  & 2 & 1.564\\
43 & 2018.09.19 & SclC16 & 0101.B-0189 & 00 58 34.1 ~~ -33 36 00.0 & 1.01 & 0.43 & 3600 & 98  & 94 & 1.686 \\\hline
44 & 2018.12.15 & Sclouter & 0102.B-0786 & 00 57 34.9 ~~ -33 53 50.0 & 1.02& 0.64  & 3600 & 39  & 37 & 0.821 
\label{table1}
\end{longtable}

\newpage

\renewcommand{\thesection}{B}
\section{Tests of the reliability of the measured \vlos\ and EWs with VLT/FLAMES LR8}
\label{app:lr8}
\setcounter{figure}{0}
\renewcommand{\thefigure}{B.\arabic{figure}} 

One observing programme in our collection (593.D-0309, PI: Battaglia) was a sequence of eight repeat observations of the same stars over 15 months, looking for velocity variability, and hence binary stars. Here to keep things simple only single pointings were compared to each other. These 
multiple VLT/FLAMES LR8 velocity and EW measurements of a sub-sample of 96 stars allow a test of the reliability (and repeatability) of velocity and EW measurements based on the CaT lines. 
We used these observations, typically with S/N$>20$, to test our error analysis, and our measurement consistency between observations made at different epochs, being careful to ignore any obviously \vlos variable stars (of which very few appear within our velocity sensitivity). A full analysis is in preparation by Arroyo Polonio et al. The fraction of binaries that they found is consistent with those expected from previous surveys.

Even though all the fields have the same setup, with the same guide and fiducial stars, and were mostly taken in good conditions, as is required for a monitoring programme, it is clear that some VLT/FLAMES observations are better behaved than others. Here we show an overview for a range of comparison plots. When the scatter in the S/N versus magnitude plot (Fig.~\ref{fig-sn}) is large, even at bright magnitudes, this is most likely due to variations in weather as well as a, perhaps related, poor astrometric alignment of the fibres compared to the guide stars. It is likely that there was a centring offset of the targets in the fibres, caused either by an imperfect astrometry of the guide stars -- either intrinsically or relative to the targets -- or an imperfect centring of the guide stars in their fibre bundles. This series of multiple measurements show how intrinsically hard it is to guarantee that guide and fiducial stars are centred correctly based on VLT/FLAMES pre-\gaia\ astrometry (Fig.~\ref{fig-sn}). The monitoring dataset does contain a range of atmospheric conditions, but there is no obvious correlation with seeing, although high air mass tended to produce poorer results, but S/N$>20$ means all results were reliable most of the time, so the problems were only seen in low S/N measurements, which strongly suggests that misalignment is the problem, as it affects fainter stars more. The size and variation of the velocity shifts between fields were clearly much smaller and more stable once \gdr{2} astrometry became available for both the targets and the guide stars and the fiducial stars for each observation. It is also worth noting that target acquisition at the telescope went much more smoothly with \gaia\ positions and proper motions. However, all the monitoring observations presented here were made pre-\gaia, so only later observations have benefited from \gaia\ astrometry.

The velocity errors were found to match the scatter in the velocities in repeat measurements of the same stars (Fig.~\ref{fig-velerr}). The velocity scatter did not appear to change with S/N, and it should vary with 1/(S/N). This is consistent with the velocity scatter being limited by the accuracy of the wavelength solution and not the S/N for this relatively high S/N sample. Any stars that are beyond the 3$\sigma$ scatter are potentially stars with intrinsically varying velocities, and thus likely binary systems (see Arroyo Polonio et al. in prep). The scatter was higher for the longer time baselines, but if this was due to slow variations intrinsic to a significant sub-sample of stars or a problem with the velocity stability over time it was not possible to say here. The intrinsic error on the velocity shifts applied to individual fields (as given in Table~\ref{table1}) was given by the accuracy of the sky lines as velocity zero points, and this was $\pm$1km/s for VLT/FLAMES LR8, and the observations tended to remain within this uncertainty. This correction likely dominated the offset velocity between fields, but the intrinsic measurement accuracy determines the scatter.

Likewise the EW scatter, and so the scatter in \feh\  determinations (Fig.~\ref{fig-feherr}) were also stable, and did not change significantly with S/N. Thus, the measurements of \feh\  appeared to be limited by the errors in the continuum estimation, probably coming from scattered light in the images, and not the S/N. Both the offset and the scatter give a picture of the intrinsic measurement errors.

There was also no evidence of any strong effect from variations in airmass or seeing on velocity or EW measurements.  In Fig. ~\ref{fig-sn} the top right hand panel, as can be seen from Table~\ref{table1}, was observed at an airmass of 1.4, but there was no noticeable effect on the S/N, or the reliability of the repeated velocity and EW measurements. In Fig.~\ref{fig-sn}, the last panel (B8) looks particularly bad for no obvious reason. In general the monitoring sample showed the consistency of the measurements, and thus provided an independent estimate of the measurement errors, which are found to be similar to previous investigations (e.g. ~\citealt{Battaglia08a}), and were most reliable for S/N$>20$. These plots highlight the intrinsic accuracy that can be achieved in the VLT/FLAMES LR8 spectroscopic measurements of velocity and \feh\ from night to night, both the zero point and the scatter.

\newpage
{\centering
\includegraphics[width=0.8\linewidth]{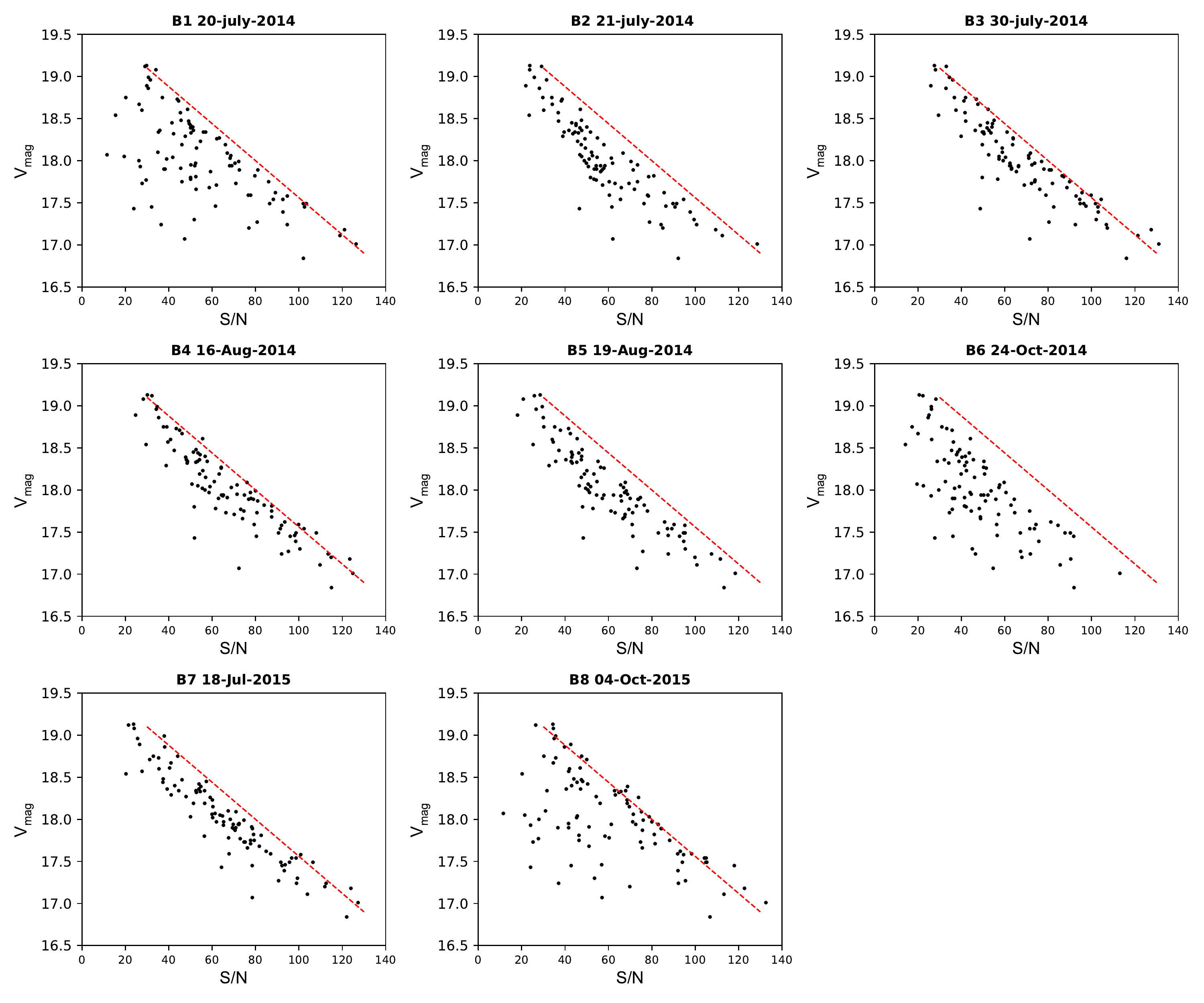}
\captionof{figure}{Variation in S/N for eight observations of the same field and the same targets with VLT/FLAMES LR8 grating. The dashed red line is always in the same position in each plot to guide the eye. The observations were made between July 2014 and October 2015 (from programme ID 593.D-0309)}
\label{fig-sn}
\vfill
\newpage
\includegraphics[width=1\linewidth]{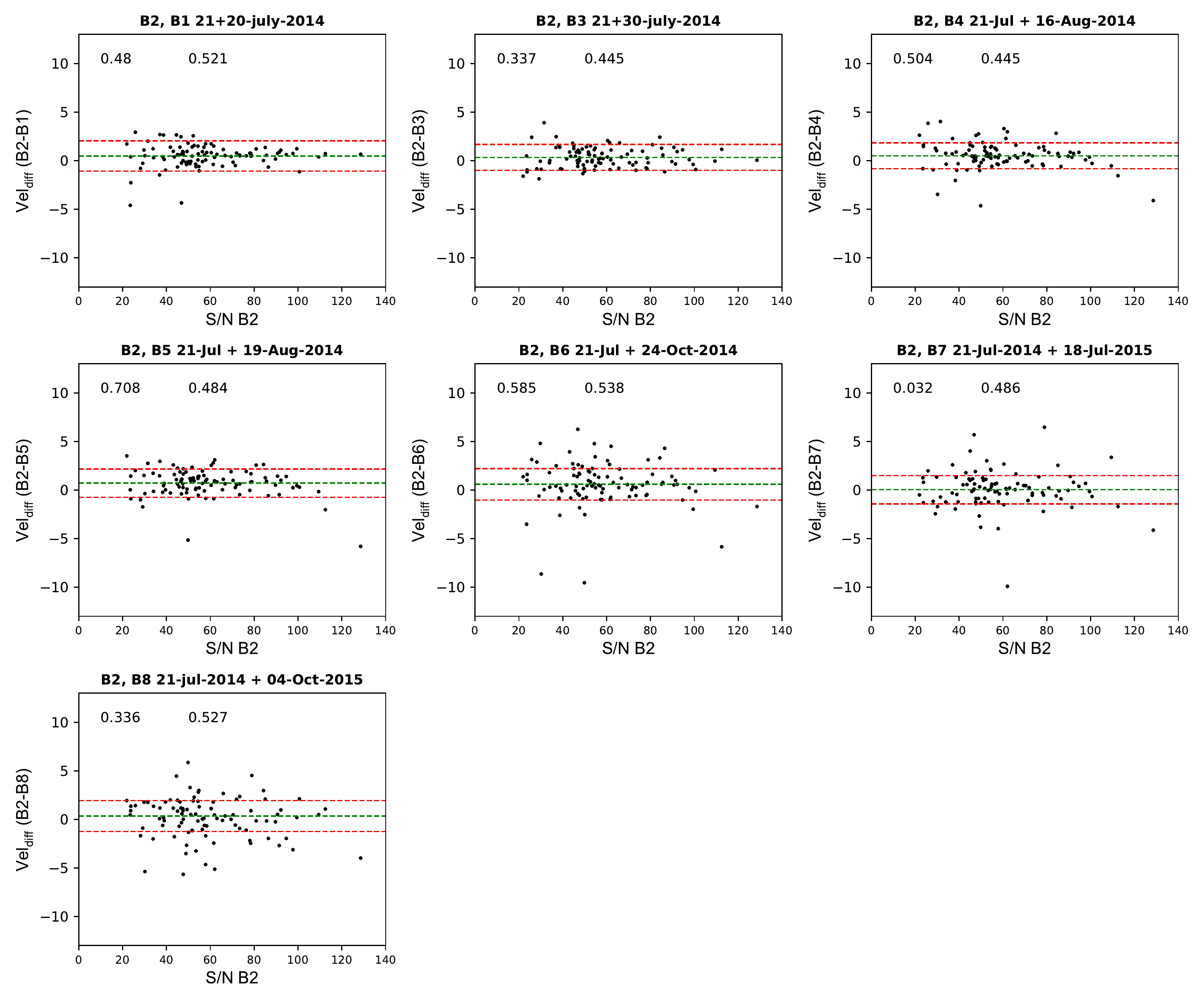}
\caption{Comparison between the velocities measured for eight VLT/FLAMES LR8 observations of the same field and the same targets as a function of the S/N of the measurement. The observations were made between July 2014 and October 2015. All velocity measurements were compared with those from observation B2, which occurred on 21 July 2014. In the top of each plot the mean velocity offset and the variance for each comparison is given; the dashed green line is the mean, and the two dashed red lines on either side show 3$\sigma$ about this mean. }
\label{fig-velerr}
\newpage 
\includegraphics[width=1\linewidth]{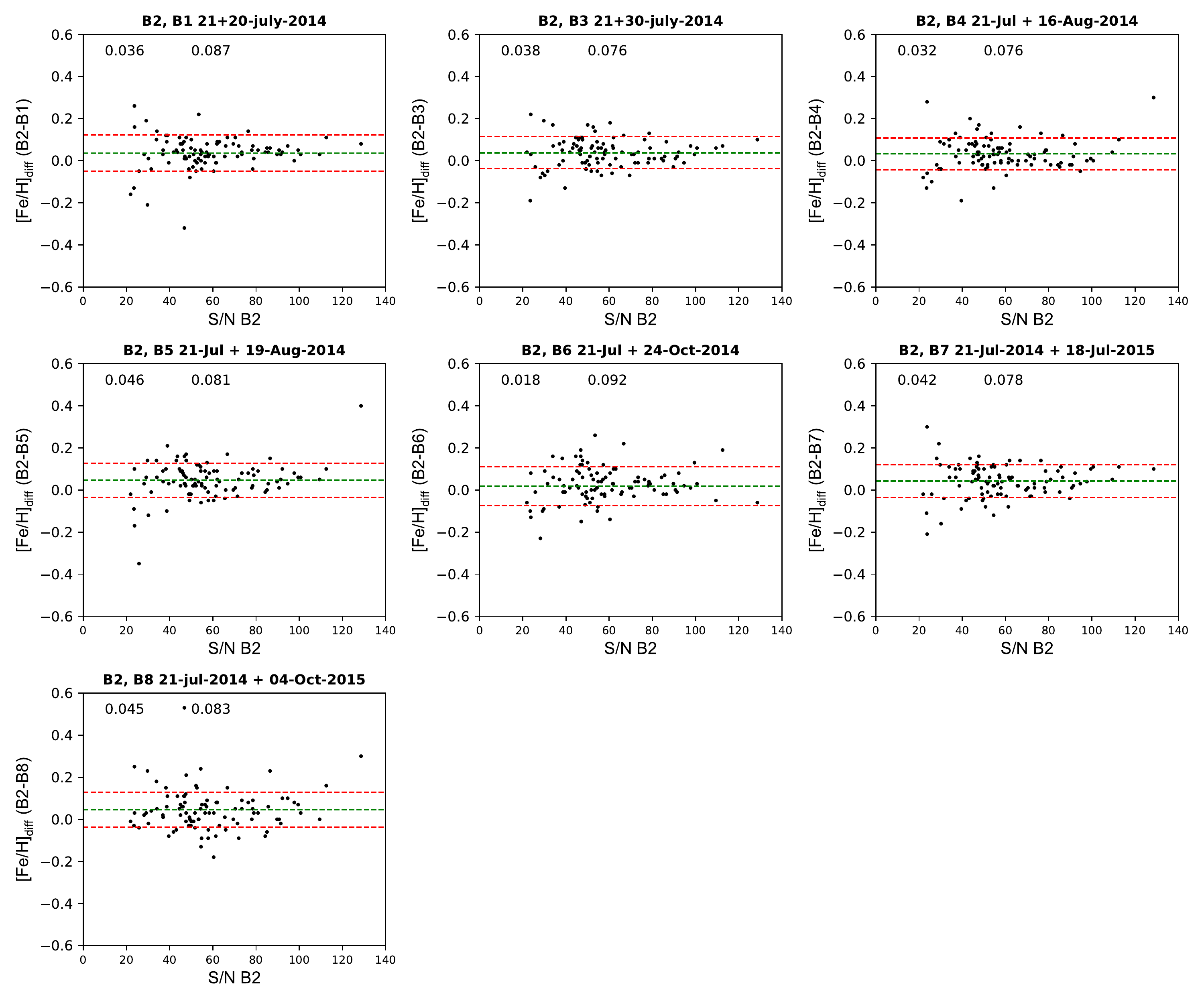}
\caption{Comparison between \feh\  determined from EW measurements of the two strongest CaT absorption lines for eight VLT/FLAMES LR8 observations of the same field and the same targets as a function of the S/N of the measurement. The observations were made between July 2014 and October 2015. All \feh\  measurements were compared with those from observation B2, which took place on 21 July 2014. In the top of each plot the mean \feh\  offset and the variance is given for each comparison; the dashed green line is the mean, and the two dashed red lines on either side show  1$\sigma$ about this mean.}
\label{fig-feherr}
}
\newpage

\renewcommand{\thesection}{C}
\section{A comparison between CaT \feh\  and direct high resolution [FeI/H] measurements}
\label{app:uves}
\setcounter{figure}{0}
\renewcommand{\thefigure}{C.\arabic{figure}} 

A subsample of the metallicities (\feh) determined using the CaT triplet lines in this paper, were compared with the VLT/FLAMES HR measurements, which were based on numerous individual Fe~I and Fe~II lines \citep[see our Fig.~\ref{uves}]{Hill19}. We used \gaia\ photometry converted to the V magnitude \citep{Riello21} applied to the metallicity conversion from EWs into \feh\ coming from \citep{Starkenburg10}. This is not the photometry that was originally assumed, so this may be the origin of a small offset of $0.1$~dex between \feh\ measured from HR10 and that from LR8, such that LR8 metallicities were 0.1~dex more metal rich than the HR measurements. There was no trend in the scatter with the magnitude of the star, nor with the changing [Ca/H], so the scatter may simply be that which could be expected due to the calibration and inherent uncertainties in the different methods.

In Fig.~\ref{uves}b we show the radial velocities measured with VLT/FLAMES HR10 compared to those measured with VLT/FLAMES LR8. The comparison looks very good, without any need for offset. There are a few outlying points, which might suggest velocity variability with time.

In Fig.~\ref{uvescat} we show how the comparison between the different methods of measuring \feh\ depend on the luminosity of the star (G mag) and also the [Ca/H] measured by \cite{Hill19}. There does not seem to be a strong trend with either magnitude or [Ca/H]. However, there are patterns, with scatter, that may suggest the known effect that the CaT determination of \feh\  also depends weakly on [Ca/H] (\cite{Starkenburg10}). 

{\centering
\includegraphics[width=1\linewidth]{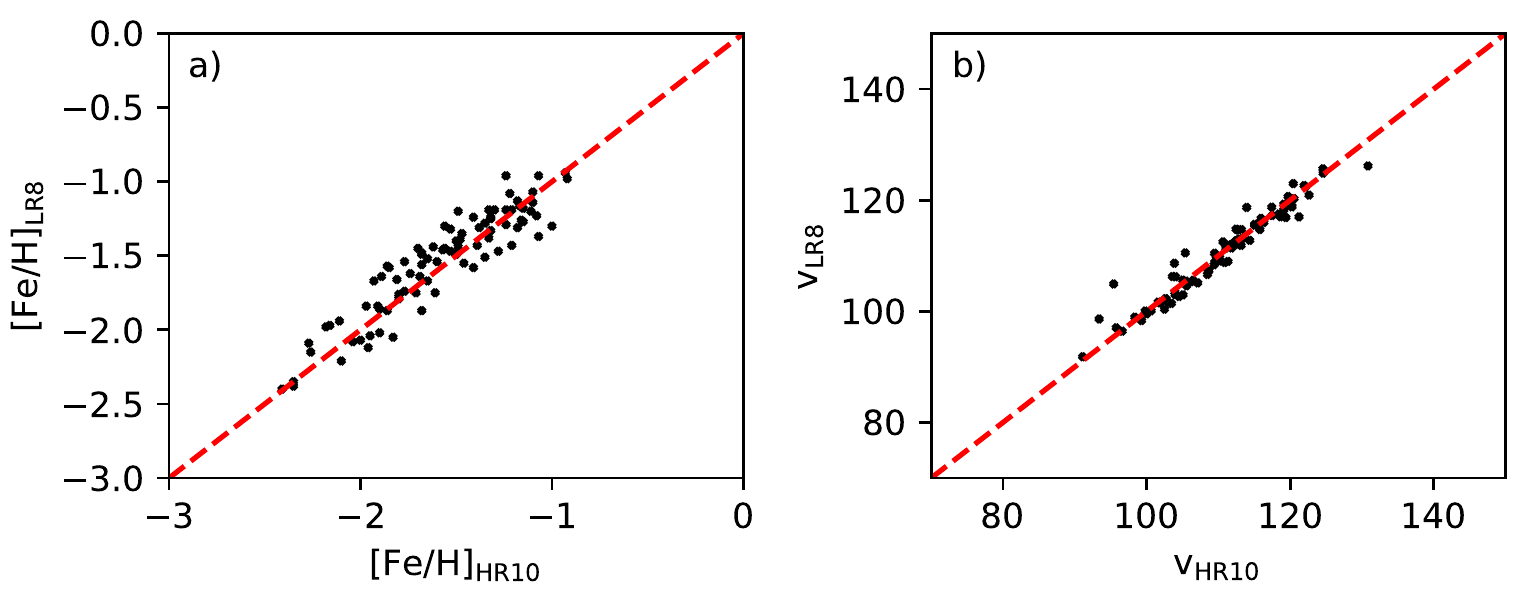}
\captionof{figure}{Comparison between VLT/FLAMES HR10 measurements and VLT/FLAMES LR8 measurements of (a)~\feh\  and (b)~the los radial velocities. }
\label{uves}
}

{\centering
\includegraphics[width=0.6\linewidth]{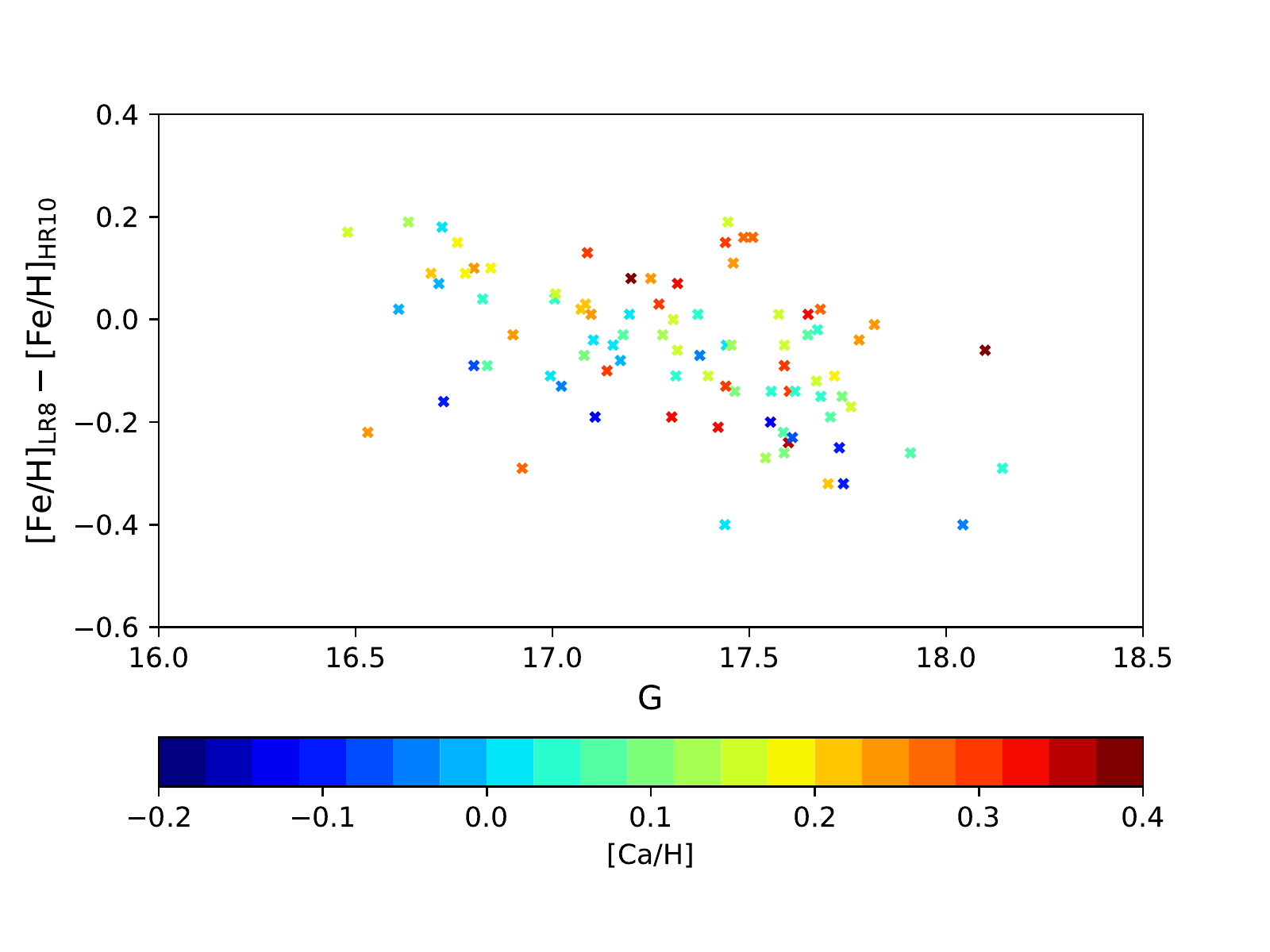}
\captionof{figure}{Comparison between VLT/FLAMES HR10 measurements and VLT/FLAMES LR8 measurements of \feh, colour coded by [Ca/H] from  VLT/FLAMES HR abundance determinations.}
\label{uvescat}
}

\newpage
\renewcommand{\thesection}{D}
\section{\gaia\ archive query for the Sculptor field}
\label{app:query}



Here we give, in Astronomical Data Query Language. the query to be executed in the \gaia\ archive, that extracts all sources in the Sculptor field with parallaxes, proper motions, and a \bpminrp\ colour. The query also retrieves the distance estimates from \cite{BJ21} and contains an on-the-fly calculation of the corrected flux excess factor $C^*$ \citep{Riello21}. The J2000.0 ICRS position and proper motion of Sculptor are as listed in the Set of Identifications, Measurements, and Bibliography for Astronomical Data \citep[SIMBAD;][]{Wenger2000} and are propagated in the query to J2016.0, the reference epoch for \gdr{3}. The selected field is two degrees in radius around the Sculptor centre.


    \begin{lstlisting}[language=SQL]
select gaia.*,
if_then_else(
    gaia.bp_rp > -20,
    to_real(case_condition(
        gaia.phot_bp_rp_excess_factor - (1.162004 + 0.011464*gaia.bp_rp 
                                    + 0.049255*power(gaia.bp_rp,2) 
                                    - 0.005879*power(gaia.bp_rp,3)),
        gaia.bp_rp < 0.5,
        gaia.phot_bp_rp_excess_factor - (1.154360 + 0.033772*gaia.bp_rp 
                                             + 0.032277*power(gaia.bp_rp,2)),
        gaia.bp_rp >= 4.0,
        gaia.phot_bp_rp_excess_factor - (1.057572 + 0.140537*gaia.bp_rp)
    )),
    gaia.phot_bp_rp_excess_factor
) as phot_bp_rp_excess_factor_corr,
cbj.r_med_geo, cbj.r_lo_geo, cbj.r_hi_geo, cbj.r_med_photogeo, 
cbj.r_lo_photogeo, cbj.r_hi_photogeo
from gaiadr3.gaia_source as gaia
join external.gaiaedr3_distance as cbj 
using(source_id)
where 
contains(
	point('ICRS',gaia.ra,gaia.dec),
	circle(
		'ICRS',
		COORD1(EPOCH_PROP_POS(15.039169999999999,-33.70889,0,
                                  .0900,.0200,111.4000,2000,2016.0)),
		COORD2(EPOCH_PROP_POS(15.039169999999999,-33.70889,0,
                                  .0900,.0200,111.4000,2000,2016.0)),
		2)
)=1 and
gaia.astrometric_params_solved>3 and gaia.bp_rp is not null
    \end{lstlisting}

\newpage
\begin{landscape}
\renewcommand{\thesection}{E}
\section{The Results}
\begin{flushleft}
Here we show an example of what is available as an online table at CDS for the combined \gdr{3} photometry and astrometry and VLT/FLAMES LR8 spectroscopy for stars in and around the Sculptor dSph.  In Table E.1 we show the full VLT/FLAMES LR8 sample, and the combination with \gdr{3} astrometry, including those likely velocity members that are in fact non-members. In Table E.2 we also include the VLT/FLAMES LR8 velocity non-members as a matter of interest, including only the measured velocities and their \gdr{3} IDs.\\
\end{flushleft}

\setcounter{table}{0}
\renewcommand{\thetable}{E.\arabic{table}}

\setlength\LTleft{0pt}            
\setlength\LTright{0pt}           


\begin{longtable}{lccccccccccccccccccccccccc}
\caption{\label{tab:all} \gdr{3} and VLT/FLAMES results for stars in and around the Sculptor dSph.}\\
\endfirsthead
\hline
\\
GaiaDR3Id      &      Gaia-RA      &   Gaia-DEC     &    Gaia-G & Gaia-BP & Gaia-RP  & plx   &  plx-err  & pmra  &  pmra-err &  pmdec  &  pmdec-err & ellr  \\
Flamesfield  &    SpecName   & N-Spec & SN  &  Vel &  vel-err & vel-var  & FeH & feh-err &  feh-var &   z &  mem & feh-rel\\
\\
\hline\hline
\\
5003449261308998016  & 13.6648395642  & -33.5229968216  & 18.122  &  18.6056 &  17.478 &   0.0258  &  0.1224 &  0.094  &  0.12   &  -0.224  &  0.167  &  1.1596 \\ 2007.gb202a  &    scl011\_09\_   & 2 &   52.97  &   119.59 &  0.418  & 0.455   &  -2.46  &  0.065 &   0.074  &   0.162 &m & 0\\ \\

5003436483781148928  & 13.7546852173  & -33.5733720453  & 19.875  &  20.3454   &19.2269 &  -0.0662 &  0.3705  & 0.728  &  0.459  &  -0.795  &  0.597  &  1.0788 \\ 
2007.gb202a   &   scl034\_11\_  &  2  &  17.02  &   105.59  & 1.49  &  3.209   &  -2.15  &  0.235  &  0.288   &  5.459 & m&  0 \\\\

5003383977805922560  & 13.8745180848 &  -33.7683070902 &  18.2596 &  18.7467  & 17.5758 &  0.153  &   0.147  &  0.008   & 0.126  &  -0.279  &  0.186  &  0.9977\\
2003.Scl11a  &    scl\_11\_040   & 3   & 25.33   &  119.23 &  0.964 &  3.043  &   -2.74 &   0.172  &  0.169  &   1.389 & m& 0 \\\\

.\\
.\\
.\\
.\\

\\
\\

\end{longtable}

\begin{flushleft}
The columns are: 
GaiaDR3Id, \gdr{3} Id number; Gaia-RA, RA (deg); Gaia-DEC, Dec. (deg); Gaia-G, \gdr{3} G magnitude; Gaia-BP, \gdr{3} G$_{BP}$ magnitude;  
Gaia-RP, \gdr{3} G$_{RP}$ magnitude;  plx, \gdr{3} parallax; plx-err, \gdr{3} error on the parallax;   pmra, \gdr{3} proper motion in RA direction;    pmra-err, \gdr{3} error on RA proper motion;  pmdec, \gdr{3} proper motion in Dec.  direction;    pmdec-err, \gdr{3} error on Dec. proper motion;  ellr, elliptical radius at which the star lies;   Flamesfield, name of the VLT/FLAMES pointing;      SpecName, name of object in VLT/FLAMES pointing;    N-Spec, number of individual VLT/FLAMES spectra combined; SN, the signal-to-noise of the single (for N-Spec=1) or combined VLT/FLAMES spectra; Vel, the \vlos for this star measured from VLT/FLAMES spectra, averaged if N-spec$>1$;  vel-err, error on the \vlos VLT/FLAMES measurement; vel-var, the variance of the VLT/FLAMES \vlos if N-spec$>1$;  FeH, the \feh\ for the star determined from the CaT measurement from VLT/FLAMES spectra, averaged if N-spec$>1$ ;  feh-err, the error on the VLT/FLAMES \feh\ determination;   feh-var,   the variance of the VLT/FLAMES \feh\ if N-spec$>1$; z, the \gdr{3} membership score, as defined in Sect. 3.1; mem, the likely membership of the star in the Sculptor dSph where m is for member stars, b is for borderline members and n is for non-members; feh-rel gives the reliablity of the \feh\ determination, where 0 is for reliable \feh\ and 1 is for an unreliable \feh, which should be used only with great caution, for non-member stars a CaT \feh\ has no meaning, and so these are given as -- in this column and there are also no values given in the table.
\vskip 1cm
The non-members in Table~\ref{tab:all} are stars with VLT/FLAMES spectroscopic \vlos\ compatible with membership in Sculptor but the \gaia\ proper motions and/or parallaxes ruled this out.

\end{flushleft}

\end{landscape}

\newpage
\renewcommand{\thetable}{E.\arabic{table}}

\begin{longtable}{lccccccc}
\caption{\label{tab:non} VLT/FLAMES spectroscopic non-members of the Sculptor dSph.}\\
\endfirsthead
\hline
\\

GaiaDR3Id       &         RA    &         Dec       &         Flamesfield    & SpecName   &   vmag  &   sn      &     vel   \\
\\
\hline\hline
\\

5026428298415447808  &   15.91929167  & -33.79244444  &           2004.Scl14a   & scl\_14\_050  &  18.31  &  25.686    &    25.26   \\

5027198781189387520   &  15.80508333  & -33.51108333    &  2004.Scl14b    & scl\_14\_064 &   19.23  &  19.638      &   63.48   \\

.\\
.\\
.\\
.\\

\\
\\

\end{longtable}

\noindent The stars in Table~\ref{tab:non} were immediately identified as \vlos\ non-members by the VLT/FLAMES spectroscopy and removed from any further analysis.

\end{appendix}

\end{document}